\RequirePackage{fix-cm}
\documentclass[smallextended]{svjour3}       
\smartqed  
\usepackage{graphicx}
\usepackage{hyperref}
 \usepackage{mathptmx}      
 \usepackage{bm}
\usepackage{graphicx}
\usepackage{amsmath}
\usepackage{amsfonts}
\usepackage{geometry}
\usepackage[dvipsnames,table]{xcolor}
\usepackage{multirow}
\usepackage[normalem]{ulem} 

\begin{document}

\title{Learning Equations from Biological Data with Limited Time Samples\thanks{This material was based upon work partially supported by the National Science Foundation under Grant DMS-1638521 to the Statistical and Applied Mathematical Sciences Institute and IOS-1838314 to KBF, and in part by National Institute of Aging grant R21AG059099 to KBF. Any opinions, findings, and conclusions or recommendations expressed in this material are those of the authors and do not necessarily reflect the views of the National Science Foundation. BM gratefully acknowledges PhD studentship funding from the UK EPSRC (reference EP/N50970X/1). AHD, LC, and KRS gratefully acknowledge funding through the NIH U01CA220378 and the James S. McDonnell Foundation 220020264. }
}


\author{John T. Nardini
\and
        John H. Lagergren
\and
        Andrea Hawkins-Daarud
\and
        Lee Curtin
\and
        Bethan Morris
\and
        Erica M. Rutter
\and
        Kristin R. Swanson
\and
        Kevin B. Flores
}


\institute{John T. Nardini, John H. Lagergren, Kevin B. Flores \at Department of Mathematics, North Carolina State University, Raleigh, North Carolina, U.S.A.
\and
John T. Nardini \at
The Statistical and Applied Mathematical Sciences Insitute, Durham, North Carolina, U.S.A.
\and
Andrea Hawkins-Daarud, Lee Curtin, Kristin R. Swanson \at
Mathematical NeuroOncology Laboratory, Precision Neurotherapeutics Innovation Program, Mayo Clinic, Phoenix, Arizona, U.S.A.
\and
Bethan Morris \at Centre for Mathematical Medicine and Biology, University of Nottingham, Nottingham, U.K.
\and
Erica M. Rutter \at Department of Applied Mathematics, University of California, Merced, Merced, California, U.S.A.
}

\date{Received: \today / Accepted: date}

\maketitle

\begin{abstract}
Equation learning methods present a promising tool to aid scientists in the modeling process for biological data. Previous equation learning studies have demonstrated that these methods can infer models from rich datasets, however, the performance of these methods in the presence of common challenges from biological data has not been thoroughly explored. We present an equation learning methodology comprised of data denoising, equation learning, model selection and post-processing steps that infers a dynamical systems model from noisy spatiotemporal data. The performance of this methodology is thoroughly investigated in the face of several common challenges presented by biological data, namely, sparse data sampling, large noise levels, and heterogeneity between datasets. We find that this methodology can accurately infer the correct underlying equation and predict unobserved system dynamics from a small number of time samples when the data is sampled over a time interval exhibiting both linear and nonlinear dynamics. Our findings suggest that equation learning methods can be used for model discovery and selection in many areas of biology when an informative dataset is used. We focus on glioblastoma multiforme modeling as a case study in this work to highlight how these results are informative for data-driven modeling-based tumor invasion predictions.
\end{abstract}

\section{Introduction}

Mathematical models are a crucial tool for inferring the mechanics underlying a scientific system of study \cite{nardini_modeling_2016} or predicting future outcomes \cite{ferguson_NPIs_2020}. The task of interpreting biological data in particular benefits from mathematical modeling, as models allow biologists to test multiple hypotheses \emph{in silico} \cite{ozik_high-throughput_2018}, optimally design experiments \cite{walter_qualitative_1990}, or create personalized medical treatment plans for patients \cite{baldock_patient-specific_2014}.  A common question for mathematicians and biologists alike is: \emph{which model(s) sufficiently describe a given data set} \cite{warne_using_2019}? This is a challenging question to resolve, as there may be several candidate models that can describe the data comparably well, or the underlying mechanics may be poorly understood. This challenge of inferring a data-driven mathematical model is further complicated by common issues with biological data that inhibit our understanding of the underlying dynamics. Such challenges include large amounts of noise \cite{francis_quantifying_2003,perretti_model-free_2013}, sparse time sampling \cite{baldock_patient-specific_2014,massey_image-based_2020,hawkins-daarud_quantifying_2019}, inter-population heterogeneity \cite{wang_prognostic_2009}, and complex forms of observation noise \cite{banks2014modeling,lagergren_learning_2020} 

Partial differential equations (PDEs) are used to model many spatiotemporal phenomena, ranging from the signaling proteins inside of cells \cite{mori_wave-pinning_2008} to the migration patterns of animal populations \cite{garcia-ramos_evolutionary_2002}. Reaction-diffusion-advection equations in particular are suited to describe biological processes that involve the simultaneous transport and growth or decay of a substance over time. Such equations may be written as
\begin{equation}
    u_t = \nabla \cdot \left (D(u,x,t) \nabla u \right) - \nabla \cdot \left( V(u,x,t) u\right) + f(u,x,t), \label{eq:RD}
\end{equation}
with system-specific initial and boundary conditions for some quantity of interest $u=u(x,t),x\in\mathbb{R}^n,t\in[t_0,t_f]$ that spreads with diffusion rate $D(u,x,t)$, migrates with advection rate $V(u,x,t)$, and grows or decays with reaction rate $f(u,x,t)$. One common example used for modeling biological phenomena is the Fisher-Kolmogorov-Petrovsky-Piskunov (Fisher-KPP) equation \cite{fisher_wave_1937}:
\begin{equation}
    u_t = D\Delta u + r u\left(1-\dfrac{u}{K}\right).\label{eq:FKPP}
\end{equation}
This equation assumes $u$ has a constant rate of diffusion $D(u,x,t)=D \in \mathbb{R}$, grows logistically with intrinsic growth rate $r\in\mathbb{R}$ until it reaches the population carrying capacity $K\in\mathbb{R}$, and does not advect ($V(u,x,t)=0$). Given an initial condition with compact support, solutions to the Fisher-KPP equation converge to traveling wave solutions that spread with constant speed $2\sqrt{Dr}$ \cite{kolmogoroff_etude_1937}. One can also show that the steepness of the propagating front depends on the ratio $D/r$; the front becomes more steep as $D/r$ decreases and becomes more spread out as $D/r$ increases, see \cite{murray_mathematical_2002} for details. 

The Fisher-KPP Equation was initially proposed to model the spread of an advantageous gene \cite{fisher_wave_1937}, but has since been used to model many biological phenomena including wound healing experiments \cite{jin_reproducibility_2016,nardini_investigation_2018,warne_using_2019} and the spread of invasive species \cite{urban_toad_2008}. In particular, it has been shown to be an informative model for glioblastoma multiforme (GBM), an aggressive brain tumor known for its heterogeneous behavior between patients \cite{baldock_patient-specific_2014,massey_image-based_2020,rockne_patient-specific_2015}. Using two time samples of standard clinical imaging in conjunction with key assumptions about how the imaging relates to cell density, the parameters $D$ and $r$ can be estimated for patients based on the wave front velocity and steepness \cite{hawkins-daarud_quantifying_2019}. Inferring these two parameters from patient data aids in characterizing the inter-patient heterogeneity common to this disease, thereby enabling data-driven modeling-based prognosis. For example, Baldock et al. \cite{baldock_patient-specific_2014} found that a patient-specific metric of invasiveness, given by $D/r$, predicts the survival benefit of gross total resection for GBM patients and, more recently, Massey et al. \cite{massey_image-based_2020} correlated this metric with temozolomide efficacy,  a  chemotherapy  drug  currently  used  in the standard of care to treat GBM.  The accurate selection of a mathematical model to aid in interpreting patient GBM data is a vital tool in informing our understanding of the efficacy of potential treatment plans for this disease.

The Fisher-KPP Equation thus serves as a useful model that provides insight into patient GBM dynamics through model prognosis and prediction of outcomes from therapeutic interventions \cite{massey_image-based_2020}. For any scientific process (GBM progression, population growth, species migration, etc.), determining which model accurately describes the underlying dynamics can be a challenging process. Common methods to choose such a model include deriving a multi-scale model from assumptions on individual interactions \cite{nardini_modeling_2016}, theorizing a heuristic model from expert knowledge, or utilizing model selection techniques \cite{bortz_model_2006,ulmer_topological_2019,warne_using_2019}. Model selection studies often consider several plausible hypotheses to describe a given process \cite{akaike_information_1998,bortz_model_2006}. Whichever of these prescribed models most parsimoniously describes the data is ultimately classified as the final selected model \cite{burnham_model_2002}. Common criteria to select models include the Akaike Information Criterion (AIC)  or the Bayesian Information Criterion \cite{akaike_new_1974,schwarz_estimating_1978}. If one has little understanding of the underlying dynamics, however, then performing a thorough model selection study may be challenging. For example, determining the form of Equation \eqref{eq:RD} that best describes a given data set can become a computationally infeasible problem due to the many possible combinations of $D(u,x,t),V(u,x,t)$, and $f(u,x,t)$ that one may need to consider. Furthermore, separate numerical methods may be needed for the accurate simulation of these different term combinations, and determining the effects of numerical error on statistical inference is an ongoing area of research \cite{nardini_influence_2019}. A robust methodology to directly infer one or a few candidate models for a given biological data set will be a valuable tool for mathematical modelers.

\emph{Equation learning} is a recent area of research utilizing methods from machine learning to infer the mathematical model underlying a given data set \cite{brunton_discovering_2016,kaiser2018sparse,lagergren_learning_2020,mangan2017model,rudy_data-driven_2017,zhang2018robust,zhang_robust_2019}. Brunton et al. introduced the Sparse Identification of Nonlinear Dynamics (SINDy) algorithm, which is able to discover the governing equations underlying the chaotic Lorenz system and other systems that can be described with ordinary differential equation models \cite{brunton_discovering_2016}. This method was extended for application to PDEs in an algorithm called PDE Functional Identification of Nonlinear Dynamics (PDE-FIND) \cite{rudy_data-driven_2017}. These studies have motivated many more investigations into how modelers can infer the governing equations for experimental data, such as how methods from Bayesian inference can be used for uncertainty quantification \cite{zhang2018robust}. Lagergren et al. recently demonstrated that a neural network can be used to reduce noise in data, which in turn improves the performance of the PDE-FIND algorithm in identifying biological transport models \cite{lagergren_learning_2020}. For example, the Lagergren et al. study found that the Fisher-KPP equation can be correctly inferred from spatiotemporal data that has been corrupted with proportional error with noise levels as high as 25\%. 

Biological data presents many challenges for equation learning methods. Equation learning studies have not been thoroughly tested in settings where data observation is limited to a small number of time samples. As many as $300$ time samples of a spatial process were observed  in inferring PDE models in \cite{lagergren_learning_2020}, and the authors of \cite{rudy_data-driven_2017} considered spatiotenporal data sets with as many as 501 time samples. Rudy et al.  \cite{rudy_data-driven_2017} demonstrated that the PDE-FIND algorithm can reliably infer model equations when only a small number of randomly-chosen spatial values are used for equation inference, but a dense number of spatiotemporal time samples were used prior to this inference step to estimate derivatives from the data. As the sampling of biological data is often sparse, equation learning methods must be robust in inferring equations with only limited time samples before they can be widely adopted for biological data.  Furthermore, many biological phenomena exhibit wide variation between realizations. To the best of our knowledge, all previous equation learning studies infer the underlying model for only a single parameter realization without considering how the final inferred equation results change over a realistic range of  parameter values. Determining how equation learning methods perform in the presence of each of these data challenges is critical if equation learning methods are to become widely adopted in the biological sciences. Such an investigation is particularly relevant for biomedical applications, including GBM growth, as clinicians may only measure the tumor volume with MR images once or twice \cite{baldock_patient-specific_2014}, and the estimated parameters for the same reaction diffusion model can vary widely between patients \cite{wang_prognostic_2009}.

We investigate two biological questions in this work: (i) how many time samples are sufficient to learn the governing equation underlying a set of data and (ii) how do these results depend on the underlying parameter values and noise levels? Such questions are of direct interest to the biological community, as data may be expensive to collect, or model parameters may not be identifiable with the available data \cite{rutter2017mathematical}. Obtaining MR images in oncology, for example, can be expensive, leading to only a small number of images for each patient. This challenge leaves GBM modelers with only one or two MR images from which they can estimate patient-specific values, including the metric of invasiveness ($D/r$) or individual diffusion and growth parameters ($D$ and $r$) \cite{baldock_patient-specific_2014,hawkins-daarud_quantifying_2019}. Similar challenges are present in ecology where an invading species' range and population size must be estimated from partial data and may only be measured annually or at a small number of locations \cite{dwyer_spatial_1998,lubina_spread_1988}. Developing methods to determine which datasets can reliably be used for equation learning is a crucial step before the broad adoption of such methods for mathematical modeling of experimental, clinical, or field data.

The goal of this work is to examine the success of equation learning methodologies in the presence of limited time sampling, large noise levels, and different parameter combinations. We focus on the Fisher-KPP Equation in this study due to its wide use in the mathematical biology literature, but the results presented in this work will extend to many common models of biological phenomena. We begin by discussing data generation and introducing our equation learning methodology in Section \ref{sec:methods}. We present our results on equation learning, fit and predicted system dynamics, parameter estimation, and uncertainty quantification in Section \ref{sec:results}. We further discuss these results in Section \ref{sec:discussion} and give final conclusions on this work and its applicability to biological studies in Section \ref{sec:conclusion}. 

\section{Methods}\label{sec:methods}

\begin{figure}
    \centering
    \includegraphics[width=0.99\textwidth]{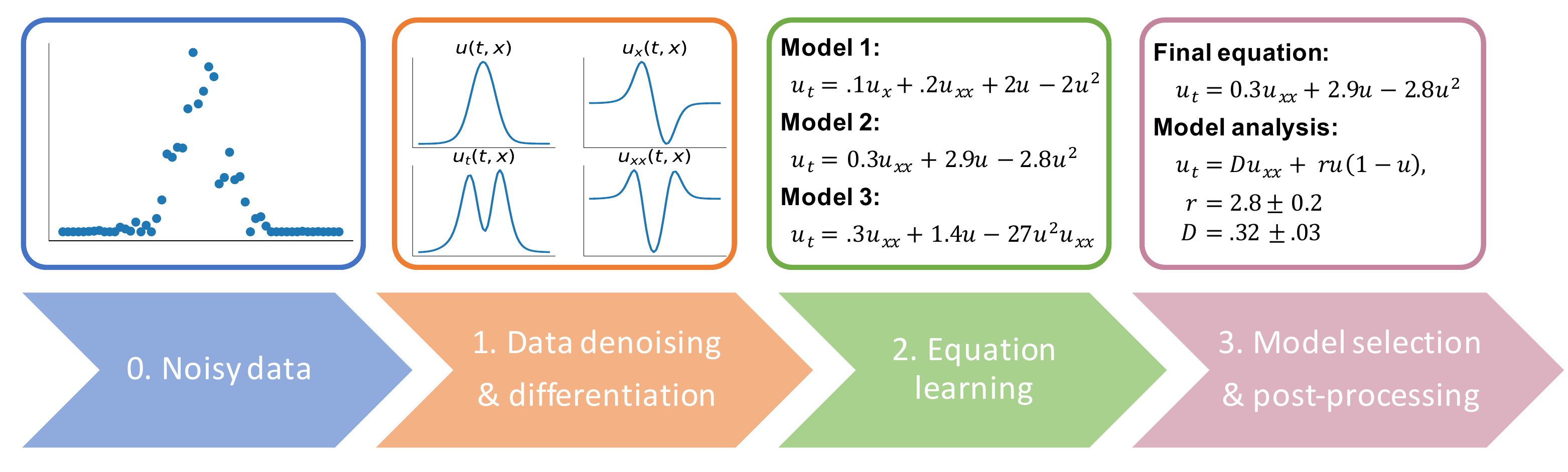}
    \caption{Visualization of our data denoising and equation learning methodology. The steps in this methodology include 1. Data denoising, where we use an ANN to smooth input noisy spatiotemporal data, which can then be used for numerical differentiation, 2. Equation learning, where we infer a small number of candidate models to describe the dynamics of the given data set, 3. Model selection and post-processing, where we infer which of these models parsimoniously describes the given data and interpret the final selected model.}
    \label{fig:EQL_pipeline}
\end{figure}

In this section we describe the data sets used throughout this study and detail the implementation of our data denoising, equation learning, and model selection methods. We discuss data generation in Section \ref{sec:data_generation}, data denoising in Section \ref{sec:data_denoising}, equation learning and model selection in Section \ref{sec:eql}, and methods for parameter estimation and uncertainty quantification from the inferred model in Section \ref{sec:PDE_Find_PE}. The equation learning methodology we adopt is summarized in Figure \ref{fig:EQL_pipeline}. The code used for this methodology is available online at \url{https://github.com/biomathlab/PDE-Learning-few-time-samples}.

\subsection{Data generation}\label{sec:data_generation}

We assume data is generated from one-dimensional spatial simulations of Equation \eqref{eq:FKPP} with an initial condition given by 

\begin{equation}
    u(x,t=0) = 0.1\text{exp}\left(-\dfrac{x^2}{.005}\right)
\end{equation}
We assume data is sampled over spatiotemporal grids of the form
\begin{align}
x_i=x_0+(i-1)\Delta x, \hspace{.5cm} i = 1,\dots,M, \hspace{.5cm} \Delta x = \frac{x_f-x_0}{M-1}  \\  
t_j = t_0 + (j-1)\Delta t,\hspace{.5cm} j = 1,\dots,N, \hspace{.5cm} \Delta t = \frac{t_f - t_0}{N-1}.
\end{align}
The spatial grid used for all simulations here is specified by $x_0=-17$ cm, $x_f=17$ cm, $M=200$. We will consider two time scales in this study by fixing $t_0=0.15$ years and letting either $t_f=0.5$ years (denoted as a \emph{short simulation}) or $t_f = 3$ years (denoted as a \emph{long simulation}) and discuss values of $N$ below.

We assume the data arise from the observation model

\begin{equation}
    y_{i,j} = u(x_i,t_j) + w_{i,j}\epsilon_{i,j};\ \ \  i = 1,\dots,M, \ \ \  j = 1,\dots,N.  \label{eq:data_generation}
\end{equation}
In Equation \eqref{eq:data_generation}, the data points, $y_{i,j}$, are assumed to be observations of Equation \eqref{eq:FKPP} that have been corrupted by noise from errors in the data collection process. Any negative values of $y_{i,j}$ are manually set to zero. The entire spatiotemporal dataset may be written as $\bm{y}=\{y_{i,j}\}_{i=1,\dots,M}^{j=1,\dots,N}$, and $\bm{y}_j$ will denote all spatial data points at the time $t_j$. To match a previous study \cite{lagergren_learning_2020}, we assume the error takes the form of a statistical model with weights given by
\begin{equation}\label{eq:prop}
    w_{i,j} = \sigma u(x_i,t_j)^\gamma.
\end{equation}
In these weights, the $\epsilon_{i,j}$ terms are independent and identically distributed (\emph{i.i.d.}) realizations of a standard normal random variable with mean 0 and variance 1, and we will set $\sigma=0.01$ or 0.05. We set $\gamma=0.5$ in this work, in which Equations \eqref{eq:data_generation} and \eqref{eq:prop} form a proportional error statistical model, meaning that the variance of the data point $y_{i,j}$ is proportional to $u(x_i,t_j)$ \cite{banks2014modeling}. We note that although we assume the constant $\gamma$ is known \emph{a priori} in this work, there exist methods for determining the value of $\gamma$ when it is unknown \cite{banks_difference-based_2016}. To further aid in the denoising and equation learning methods presented later on, we manually remove spatial data points from each dataset for which the value of $|y_{i,j}|$ never exceeds $10^{-4}$ for any time points $j=1,\dots,N$. 

\begin{figure}[ht]
    \centering
    \includegraphics[width=0.45\textwidth]{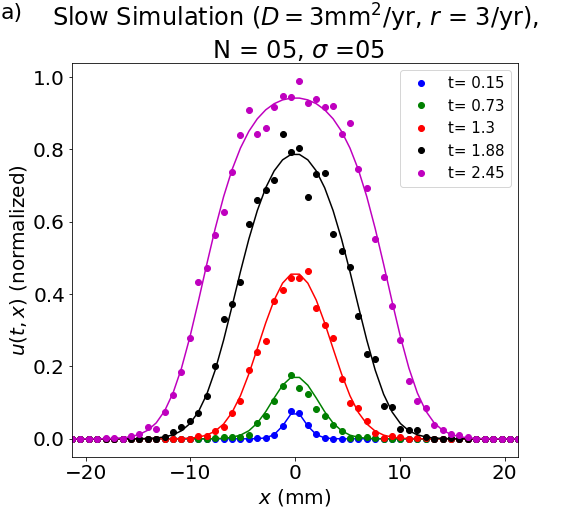}
    \includegraphics[width=0.45\textwidth]{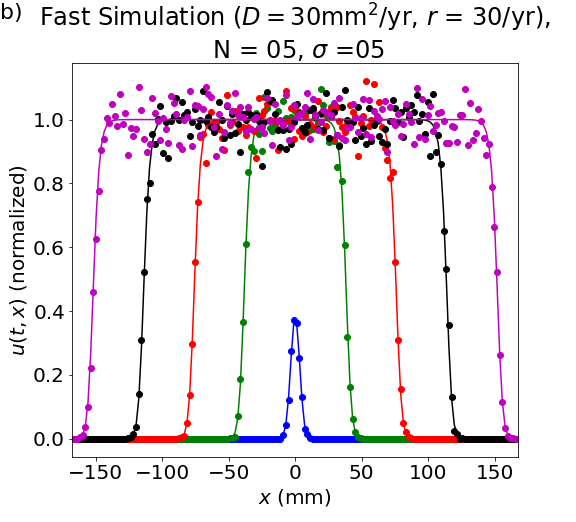}
    \includegraphics[width=0.45\textwidth]{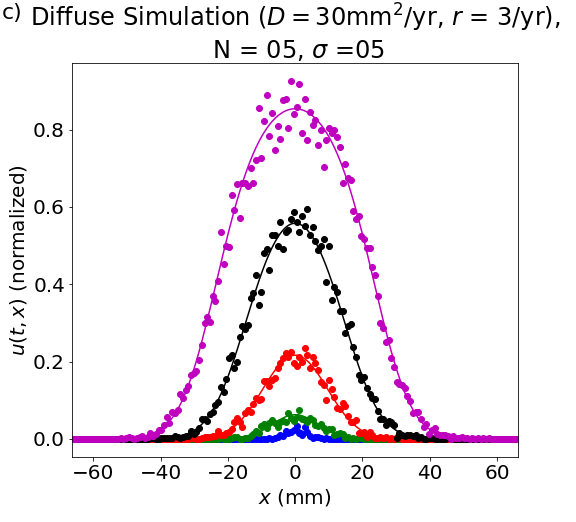}
    \includegraphics[width=0.45\textwidth]{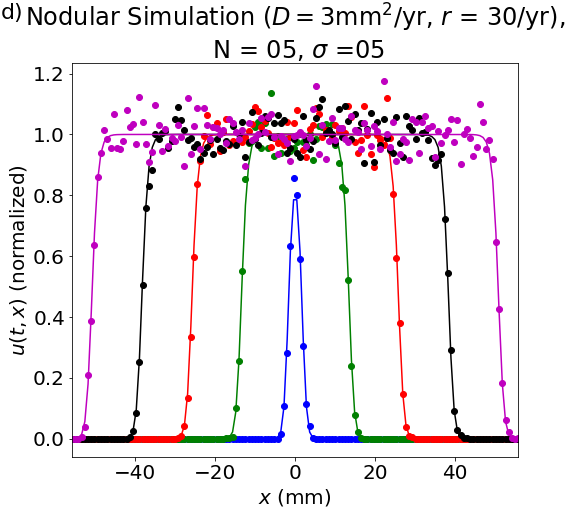}
    \caption{Simulated noisy data of the Fisher-KPP equation with 5\% noise and $N=5$ time samples for the (a) Slow Simulation, (b) Fast Simulation, (c) Diffuse Simulation, and (d) Nodular Simulation. Solid lines represent noiseless simulations of Equation \eqref{eq:FKPP} and points represent observations generated by Equation \eqref{eq:data_generation}. Blue plots correspond to $t=0.15 $ years, green plots correspond to $t=0.73$ years, red plots correspond to $t=1.3$ years, black plots correspond to $t=0.88 $ years, and magenta plots correspond to $t=2.45 $ years.}
    \label{fig:data}
\end{figure}

\begin{figure}[h]
    \centering
    \includegraphics[width=0.48\textwidth]{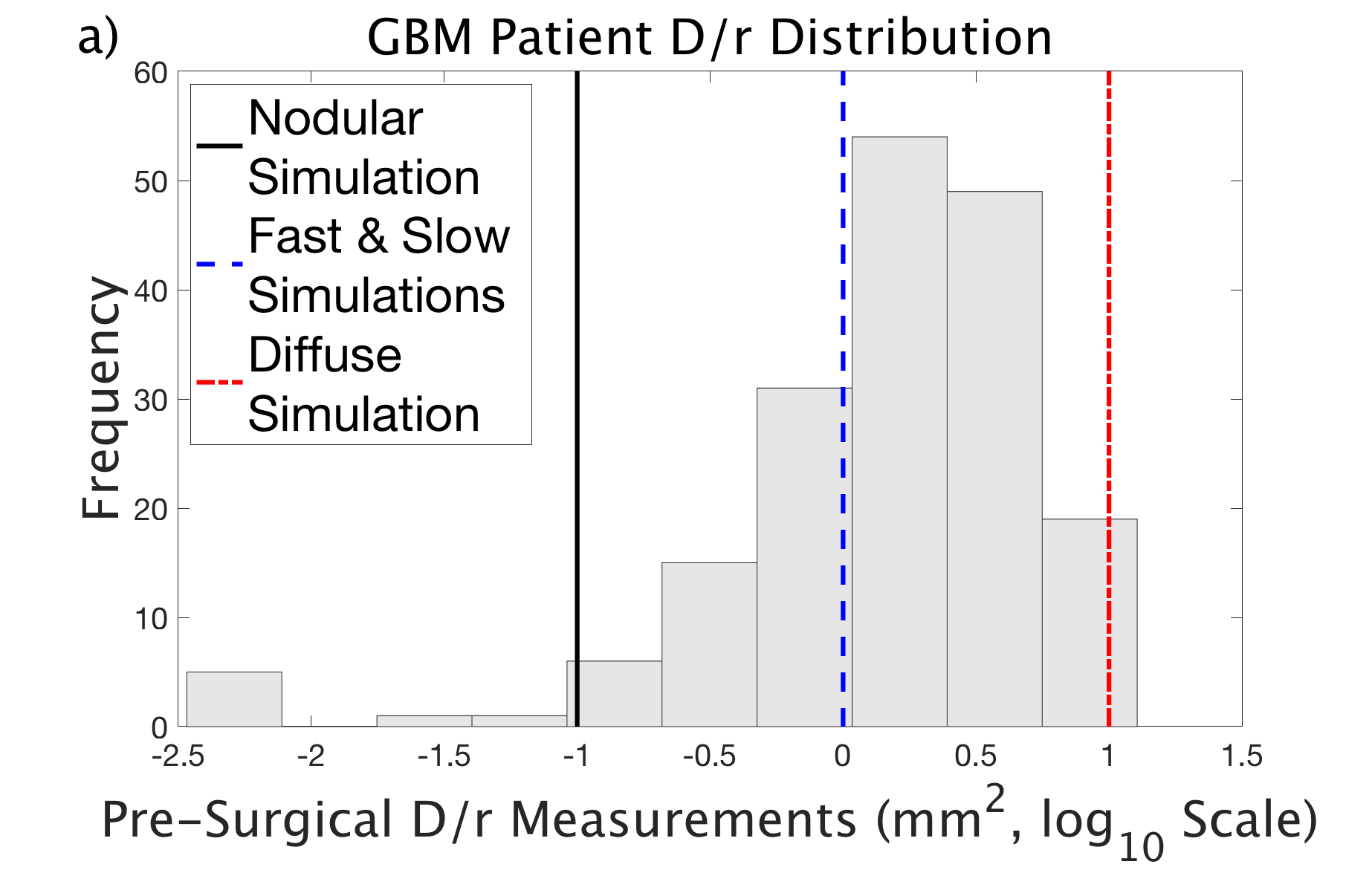}
    \includegraphics[width=0.48\textwidth]{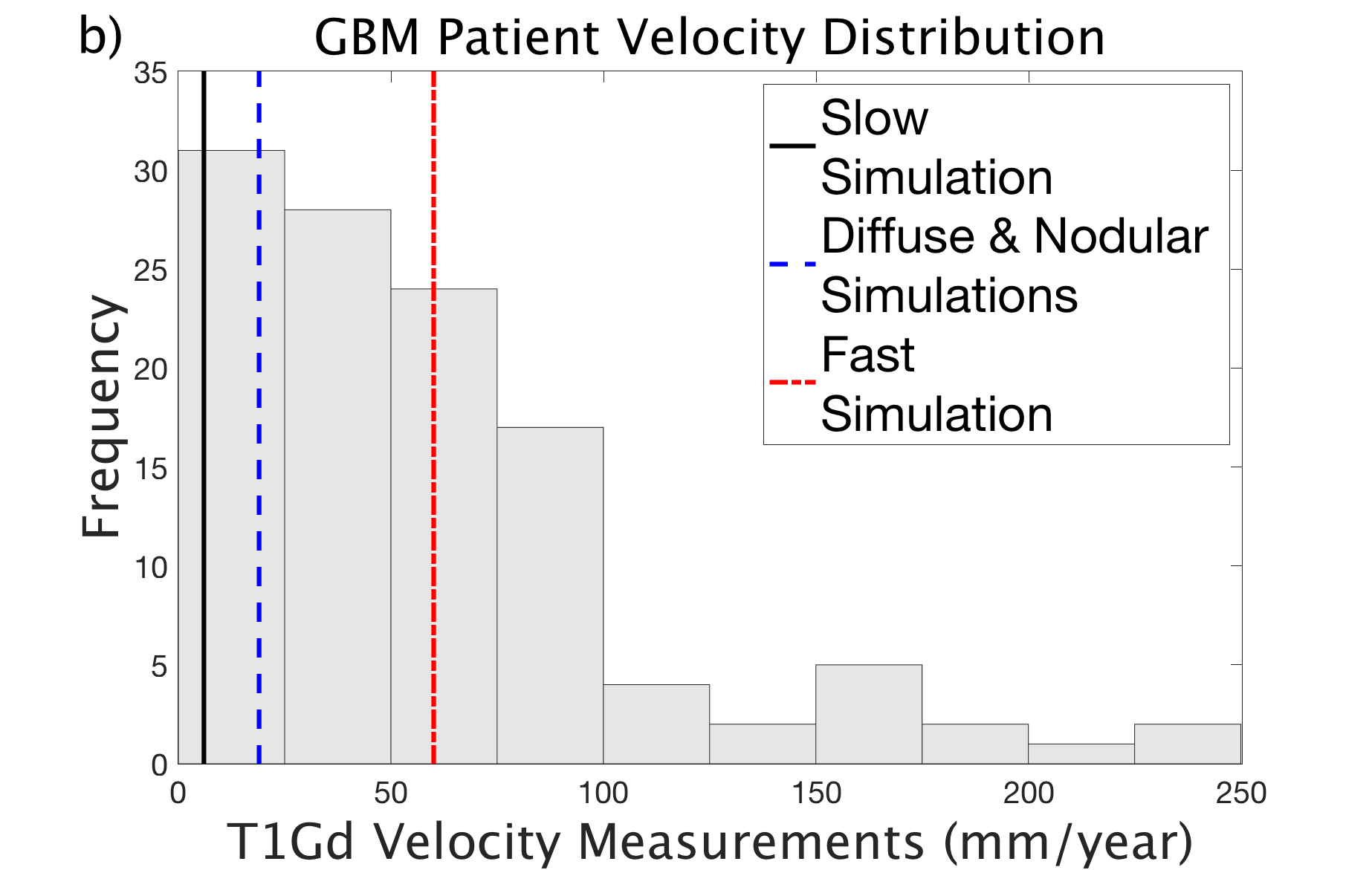}
    \caption{Histograms of measured GBM patient values from \cite{neal2013_cancerresearch,neal2013_plosone}, including (a) Pre-surgical $D/r $ values and (b) T1Gd Velocity.}
    \label{fig:histograms}
\end{figure}

We are concerned with several data aspects in this study to investigate the performance of our methodology in the presence of biological data challenges. Namely, we vary the number of time samples $N$, the parameter values $(D,r)$ parameterizing $u(x,t)$, whether data is sampled over the long or short time interval, as well as the noise level in the data. We will vary each of these values over the following domains: $N=\{3,5,10\}$, $(D,r)=\{(3,3),(30,30),(30,3),(3,30)\}$, $t_f=\{0.5,3.0\},$ and $\sigma=\{0.01,0.05\}$. The units for $D$,$r$, and $t_f$ are $\text{mm}^2/\text{year}$, $1/\text{year}$, and year, respectively. We will refer to each of these $(D,r)$ combinations as slow, fast, diffuse, and nodular simulations, respectively. Figure \ref{fig:data} depicts resulting datasets from these four simulations over the long time interval. These simulation names and parameters are borrowed from a previous GBM modeling study \cite{hawkins-daarud_quantifying_2019} to cover the ranges of $D$ and $r$ that have been inferred for GBM patients. Figure \ref{fig:histograms} presents histograms of measured metrics of invasiveness ($D/r$) and velocities from 200 GBM patients as well as where the four simulations from this study fall in these distributions \cite{baldock_patient-specific_2014,neal2013_cancerresearch,neal2013_plosone}.  Table \ref{tab:variable_ranges} summarizes the range of parameters used in this study, as well as their labels used throughout. 

\begin{table}[ht]
    \centering
    \begin{tabular}{|c|c|c|}
        \hline
        Variable & Realized Values & Labels \\
        \hline
        N & 3,5,10 & 3,5,10 \\
        $\left(D,r \right)$ & $(3,3),(30,30),(30,3),(3,30)$ & slow, fast, diffuse, nodular \\
        $t_f $ & 0.5,3 & short, long\\
        $\sigma$ & 0.01,0.05 & 1\%,5\% \\
        \hline
    \end{tabular}
    \caption{Summary of the varied parameters used throughout this study. Units for $D$ are mm$^2$/year , $r$ are 1/year , and $t_f$ are year.}
    \label{tab:variable_ranges}
\end{table}

\subsection{Data denoising}\label{sec:data_denoising}

The first step in the equation learning process is data denoising and differentiation. Data denoising for equation learning refers to the process of approximating the solution and derivatives of an underlying dynamical system from a set of observed data. Previous methods relied on the use of finite differences for numerical differentiation in the case of noiseless data and polynomial splines in the presence of noise \cite{boninsegna2018sparse,rudy_data-driven_2017,zhang2018robust}. Lagergren et al. recently introduced a method that leverages artificial neural networks (ANNs) to be used for robust data smoothing and numerical differentiation in the presence of large amounts of noise and heteroscedastic data \cite{lagergren_learning_2020}. 

 A data denoising ANN, denoted by $h(\bm{x}|\theta)$, is a differentiable function with inputs consisting of spatiotemporal points, $\bm{x}=(x,t)$, and outputs consisting of approximations for $u(\bm{x})$. The ANN is parameterized by the set $\theta$ consisting of weight matrices $W_i$ and bias vectors $b_i$ for $i=1,\dots,L+1$ where $L$ indicates the number of hidden layers or ``depth'' of the ANN. The size of each weight matrix and bias vector is determined by the number of neurons or ``width'' of the corresponding layer and its preceding layer. In particular, for the weight matrix and bias vector in the $i^{\text{th}}$ layer, the width of the $i-1$ layer determines the number of rows of the weight matrix while the width of the $i^{\text{th}}$ layer gives the number of columns and length of the weight matrix and bias vector, respectively. ``Activation functions'' (e.g. the sigmoid function $\sigma(x) = 1/(1+\text{e}^{-x})$) are applied element-wise in between layers to introduce nonlinearity to the ANN. 

\begin{figure}[ht]
    \centering
    \includegraphics[width=0.75\textwidth]{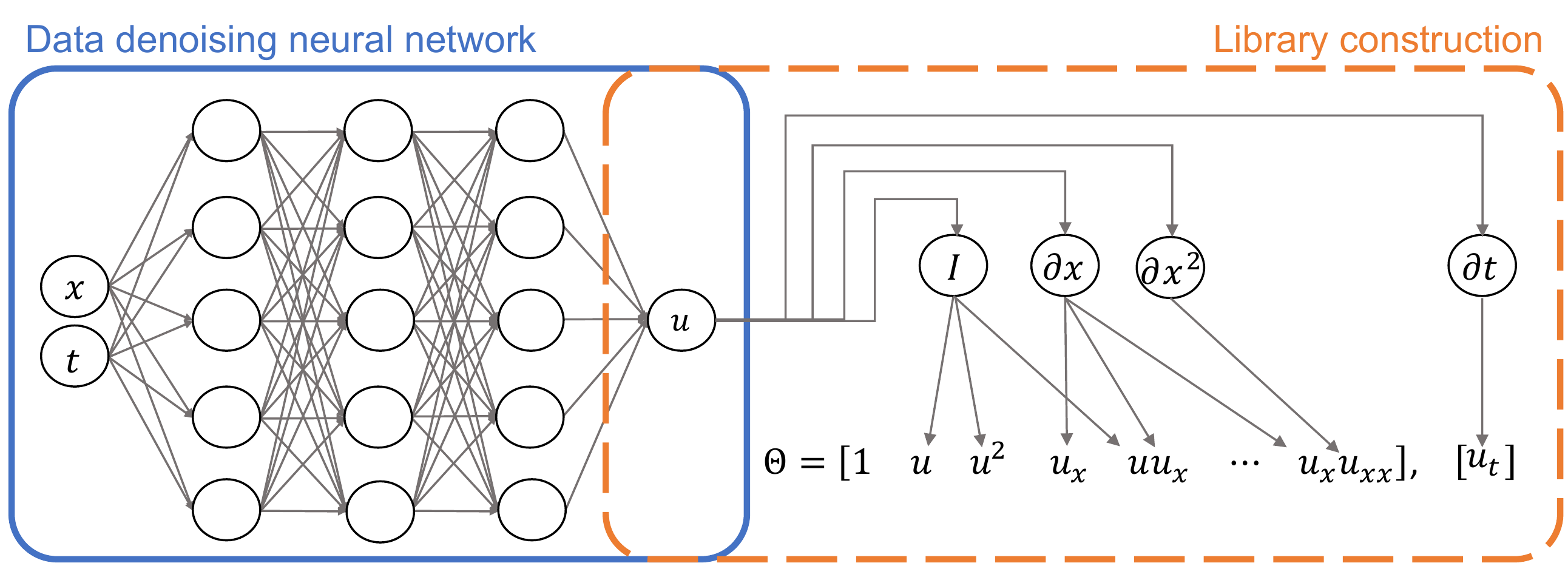}
    \caption{Diagram of the equation learning pipeline. Left: data denoising step in which an artificial neural network (ANN) is used to approximate the solution to the dynamical system. Right: library construction step where partial derivatives of the trained ANN are used to construct the left hand side ($u_t$) and library of candidate right hand side terms ($\Theta$) of the desired PDE.}
    \label{fig:pipeline}
\end{figure}

With appropriately chosen activation functions, ANNs are in the class of functions known as \emph{universal function approximators} \cite{hornik_1991}. In practice, this means that given large enough width and depth, ANNs can approximate any continuous function on a compact domain arbitrarily well. Therefore, following \cite{lagergren_learning_2020}, we use ANNs as surrogate models for the quantity of interest, $u(x,t)$. Back propagation of a trained ANN can be used for numerical differentiation, which is used to construct candidate terms for the equation learning task. See Figure \ref{fig:pipeline} for a diagram of the data denoising procedure.

The ANN in this work is chosen to have three hidden layers with 256 neurons in each layer. This choice is large enough to give the ANN sufficient capacity to approximate the solution $u$. Concretely, this ANN can be written as
\begin{equation}\label{eq:ann}
    h(\bm{x}|\theta) = \phi\Bigg(\,\sigma\Big(\,\sigma\big(\,\sigma(\,\bm{x}W_1 + b_1\,)\,W_2 + b_2\,\big)\,W_3 + b_3\,\Big)\,W_4 + b_4\,\Bigg)
\end{equation}
where the weight matrices are denoted by $W_1\in\mathbb{R}^{2\times256}$, $W_2, W_3\in\mathbb{R}^{256\times256}$, and $W_4\in\mathbb{R}^{256\times1}$, bias vectors are denoted by $b_1, b_2, b_3\in\mathbb{R}^{256}$, and $b_4\in\mathbb{R}$, and activation functions are denoted by $\sigma(x) = 1/(1+\text{e}^{-x})$ and $\phi(x) = \log(1 + \text{e}^x)$. Note that $\phi$ (also called the ``softplus'' function) is specifically chosen to constrain neural network outputs to $[0, \infty)$ since we assume values of $u(\bm{x})$ are non-negative. 

The ANN parameters $\theta$ are optimized using the first-order gradient-based Adam optimizer \cite{kingma_adam_2017} with default parameters and batch-optimization. The ANN is trained by minimizing the objective function given by:
\begin{equation}\label{eq:train_loss}
    \mathcal{L}(\theta) = \sum_{i,j}^{M,N} \left(\dfrac{ h(x_i,t_j|\theta) - y_{i,j} }{|h(x_i,t_j|\theta)|^\gamma}\right)^2 + \sum_{i,j}^{M,N} h(x_i,t_j|\theta)^2\Big|_{h(x_i,t_j|\theta)\not\in [u_{\text{min}},u_{\text{max}}]} 
\end{equation}
over the set of ANN parameters $\theta$. The first term of Equation \eqref{eq:train_loss} corresponds to the generalized least squares error between the ANN $h(\bm{x}|\theta)$ and the observed noisy data $\bm{y}$. The second term acts as a regularization penalty to ensure ANN values stay between the minimum and maximum observed values of $u$. We assume that $u$ is normalized so that $u_\text{min} = 0, u_\text{max} = 1$. Note that in the small-time-sample limit (e.g. three time samples), the ANN can output unrealistic values between time samples. Therefore, for a given batch of data, the first term of Equation \eqref{eq:train_loss} is evaluated on the corresponding batch of observed data values while the second term is evaluated on a fixed $100\times100$ grid spanning the input domain. 

To prevent the ANN from overfitting to the data, the following measures are taken. The training data are randomly partitioned into 80\%/20\% training and validation sets. The network parameters are iteratively updated to minimize the error in Equation \eqref{eq:train_loss} on the training set. The network that results in the best error on the validation set is saved. Note that data from the validation set are never used to update network parameters. We use a small batch size of 10 to train each ANN. A small batch size acts as an additional form of regularization that (i) helps the ANN escape local minima during training and (ii) allows for better generalization \cite{keskar_large-batch_2017}. We employ early stopping of 1,000 (i.e. training ends if the validation error does not decrease for 1,000 consecutive epochs) to ensure convergence of each ANN independently.

The trained ANN is then used in the following Equation Learning step to build a library of candidate equation terms consisting of denoised values of $u$ and the approximated partial derivatives of $u$. All network training and evaluation is implemented in Python 3.6.8 using the PyTorch deep learning library. 

\subsection{Equation learning, model selection, and equation post-processing}\label{sec:eql}

The second step of our equation learning methodology is to infer several candidate models to describe the data's dynamics. From the approximations of $u$ and its partial derivatives from the ANN, we use the PDE-FIND algorithm \cite{rudy_data-driven_2017} to infer data-driven governing equations. This algorithm consists of first building a large library of candidate terms for the inferred model and then selecting which of these terms to include in the final model.

The PDE-FIND algorithm builds a library of potential right hand side terms in the matrix, $\Theta$. The columns of $\Theta$ are comprised of candidate terms for the final inferred models. We use a library consisting of ten terms given by $\Theta = [u_{x}, u_{xx}, u, u^2, uu_{x}, u^2u_{x}, uu_{xx}, u^2u_{xx}, u_{x}^2, uu_{x}^2].$

From  $\Theta,$ the PDE-FIND algorithm then recovers which terms should be included in the final inferred model by solving the linear system
\begin{equation}
    u_t = \Theta\xi, \label{eq:PDE-FIND}
\end{equation}
where $\xi$ is a sparse vector parameterizing the final inferred differential equation model. Nonzero elements of $\xi$ correspond to relevant right hand side terms of the inferred model. Sparse linear regression methods are used to solve Equation \eqref{eq:PDE-FIND} sparsely to ensure that the final recovered model is simple and contains a small number of terms. Following our previous study \cite{lagergren_learning_2020}, we use the adaptive forward-backward Greedy algorithm to estimate $
\xi$ \cite{zhang2009adaptive}. This algorithm  requires estimation of the optimal tolerance with which to solve Equation \eqref{eq:PDE-FIND}; to estimate this hyperparameter, we randomly split the spatiotemporal domain into 50\%/50\% training and validation sets. Our previous study showed that small but systematic biases from the ANN can lead to incorrect terms being inferred, so we incorporate a round of ``pruning'' after solving Equation \eqref{eq:PDE-FIND}. During this round of pruning, we check the sensitivity of the mean-squared error $\|u_t-\Theta\hat{\xi}\|_2^2$ to each nonzero entry in $\xi$.  The $i^{\text{th}}$ term of $\hat{\xi}$ is included in the final inferred equation if $\|u_t-\Theta_i\hat{\xi_i}\|_2^2$ increases by 5\%, where $\Theta_i$ is a copy of $\Theta$ that is missing the $i^{\text{th}}$ term and $\hat{\xi_i}$ is the estimated parameter vector in the absence of this term. The PDE-FIND algorithm is sensitive to the training-validation split of the spatiotemporal domain, so we infer 100 equations from 100 random training-validation domains and choose the three most common forms as the final inferred equations.

The third step of our equation learning methodology is model selection and post-processing of these inferred equations. To select which of the three top models best describes a given dataset, we implement the following model selection procedure. We perform forward simulations of these three top models starting with initial conditions from the first time sample, $\bm{y}_1$. Numerical implementation of the inferred equations is discussed in Appendix \ref{sec:simulating_learned}. From these three models, we select the model with the lowest AIC score when compared to data at the final time sample $\bm{y}_N$ as our selected model. The AIC here is advantageous because when two models lead to similar outputs, it will select the simpler model and avoid complex models. We include some  post-processing of this model to ensure that the final model is interpretable. In particular, we ensure that the transport terms in the final recovered model can be written in a flux formulation and in turn conserve mass. For example, the diffusion term $u_{xx}$ by itself can be re-written as $u_{xx} = (u_x)_x$ and is thus in conservation form. The term $u_{x}^2$ by itself cannot be re-written in conservation form so it would be manually removed. If the two terms $u_{x}^2$ and $uu_{xx}$ are simultaneously recovered, then we keep both terms because $u_{x}^2 + uu_{xx} = (uu_x)_x$. 

\subsection{Parameter Estimation and Uncertainty Quantification} \label{sec:PDE_Find_PE}

We will consider the performance of the PDE-FIND algorithm in parameter estimation for a given data set by solving Equation \eqref{eq:PDE-FIND} where $\Theta$ is composed of the terms in Fisher-KPP Equation. We build the two column library of terms given by $\Theta = [u(1-u),u_{xx}]$ and solve the linear system $u_t = \Theta\xi$ using the \textbf{linalg.lstsq} function in the Python Numpy subpackage. We perform this computation for 100 separate training sets comprised of 50\% of the spatiotemporal domain to determine how much these parameter estimates vary. This method of randomly selecting several subsambles of the domain for uncertainty quantification is known as subagging (a sobriquet for subsample aggregrating) \cite{buhlmann_bagging_2012}.  We set our parameter estimate to be the median of these 100 estimates, and we set its \emph{normalized standard error} to be the standard error of these estimates divided by the median value.

\section{Results}\label{sec:results}

We investigate the performance of our equation learning methodology in the presence of sparse time sampling over both the short and long timescales. We exhibit these results using artificial data sets simulated  from the four simulations described in Table \ref{tab:variable_ranges}. Equation inference is discussed in Section \ref{sec:results_eql}, and parameter estimation and uncertainty quantification when the underlying equation is known are discussed in Section \ref{sec:results_params_UQ}.

\subsection{Equation Discovery}\label{sec:results_eql}

We test the performance of our equation learning method on artificial data that has been generated from all four simulations for N = 3, 5, or 10 time samples over both the short (0-0.5 years) and long (0-3 years) timescales with 1\% noise (Table \ref{tab:EQL_01}) and 5\% noise levels (Table \ref{tab:EQL_05} in Appendix \ref{sec:learning_1d_long_supp}). We investigate whether our equation learning methodology is able to infer the Fisher-KPP data from noisy data and how well the final inferred equations describe the true underlying dynamics. We test the model's predicted dynamics by investigating both how the inferred models match the dynamics they were trained on and how the models predict system dynamics that they were not trained on. We further consider our methodology's ability to infer the correct underlying equation form over a finer range of $(D,r)$ values in Section \ref{subsec:EQL_param_sweep}.

\begin{table}
\centering
\begin{tabular}{|c|c|c|c|}
\hline
\multicolumn{2}{|c|}{\textbf{Slow Simulation}} & \multicolumn{2}{c|}{$\bm{u_t = 3u_{xx} + 3u-3u^2}$} \\
\hline
$\sigma$ & \ $N$ & Learned Equation (0-0.5 years) & Learned Equation (0-3 years) \\ 
\hline
     01 & 03 & $u_t = -63.082u^2 + 6.394u$ & $u_t = 2.9u_{xx}-2.601u^2 + 2.723u$ \cellcolor [HTML]{77dd77} \\
\hline
     01 & 05 & $u_t = 2.0u_{xx}-22.572u^2 + 3.861u$ \cellcolor [HTML]{77dd77} & $u_t = 1.8u_{xx}-3.125u^2 + 3.061u$ \cellcolor [HTML]{77dd77} \\
\hline
     01 & 10 & $u_t = 0.8u_{xx}-44.634u^2 + 5.418u$ \cellcolor [HTML]{77dd77} & $u_t = 2.3u_{xx}-3.086u^2 + 3.044u$ \cellcolor [HTML]{77dd77} \\
\hline
\multicolumn{4}{|c|}{ } \\
\hline
\multicolumn{2}{|c|}{\textbf{Diffuse Simulation}} & \multicolumn{2}{c|}{$\bm{u_t = 30u_{xx} + 3u-3u^2}$} \\
\hline
$\sigma$ & \ $N$ & Learned Equation (0-0.5 years) & Learned Equation (0-3 years) \\ 
\hline
     01 & 03 & $u_t = -195.481u^2 + 6.583u$ & $u_t = 15.3u_{xx}-3.285u^2 + 2.485u$ \cellcolor [HTML]{77dd77} \\
\hline
     01 & 05 & $u_t = -1.1u_{x} + 31.5u_{xx} + 1.303u$ & $u_t = 21.3u_{xx}-3.483u^2 + 3.283u$ \cellcolor [HTML]{77dd77} \\
\hline
     01 & 10 & $u_t = 29.4u_{xx} + 2.815u$ & $u_t = 27.9u_{xx}-3.023u^2 + 3.003u$ \cellcolor [HTML]{77dd77}\\
\hline
\multicolumn{4}{|c|}{ } \\
\hline
\multicolumn{2}{|c|}{\textbf{Fast Simulation}} & \multicolumn{2}{c|}{$\bm{u_t = 30u_{xx} + 30u-30u^2}$} \\
\hline
$\sigma$ & \ $N$ & Learned Equation (0-0.5 years) & Learned Equation (0-3 years) \\ 
\hline
     01 & 03 & $u_t = -42.113u^2 + 43.357u + 176.8uu_{xx}-272.1u_{x}^2$ & $u_t = 62.4u_{xx}-260.1u^2u_{xx}$ \\
\hline
     01 & 05 & $u_t = 25.7u_{xx}-34.379u^2 + 34.385u$ \cellcolor [HTML]{77dd77} & $u_t = -53.1u_{x} + 69.2u_{xx} + 7.73u^2u_{x}$ \\
\hline
     01 & 10 & $u_t = 21.1u_{xx}-32.383u^2 + 32.471u$ \cellcolor [HTML]{77dd77} & $u_t = 77.9u_{xx}-274.1u^2u_{xx}$ \\
\hline
\multicolumn{4}{|c|}{ } \\
\hline
\multicolumn{2}{|c|}{\textbf{Nodular Simulation}} & \multicolumn{2}{c|}{$\bm{u_t = 3u_{xx} + 30u-30u^2}$} \\
\hline
$\sigma$ & \ $N$ & Learned Equation (0-0.5 years) & Learned Equation (0-3 years) \\ 
\hline
     01 & 03 & $u_t = 3.2u_{xx}-27.909u^2 + 28.125u$ \cellcolor [HTML]{77dd77} & $u_t = -22.315u^2 + 22.316u$ \\
\hline
     01 & 05 & $u_t = 3.0u_{xx}-30.294u^2 + 30.406u$ \cellcolor [HTML]{77dd77} & $u_t = -28.583u^2 + 28.548u$ \\
\hline
     01 & 10 & $u_t = 2.5u_{xx}-30.361u^2 + 30.31u$ \cellcolor [HTML]{77dd77} & $u_t = -30.215u^2 + 30.192u$ \\
\hline
\end{tabular}
\caption{Learned 1d Equations from our equation learning methodology for all simulations with 1\% noisy  data. Correctly-inferred equation forms are shaded in green. }
\label{tab:EQL_01}
\end{table}

\begin{figure}
    \centering
    \includegraphics[width=0.45\textwidth]{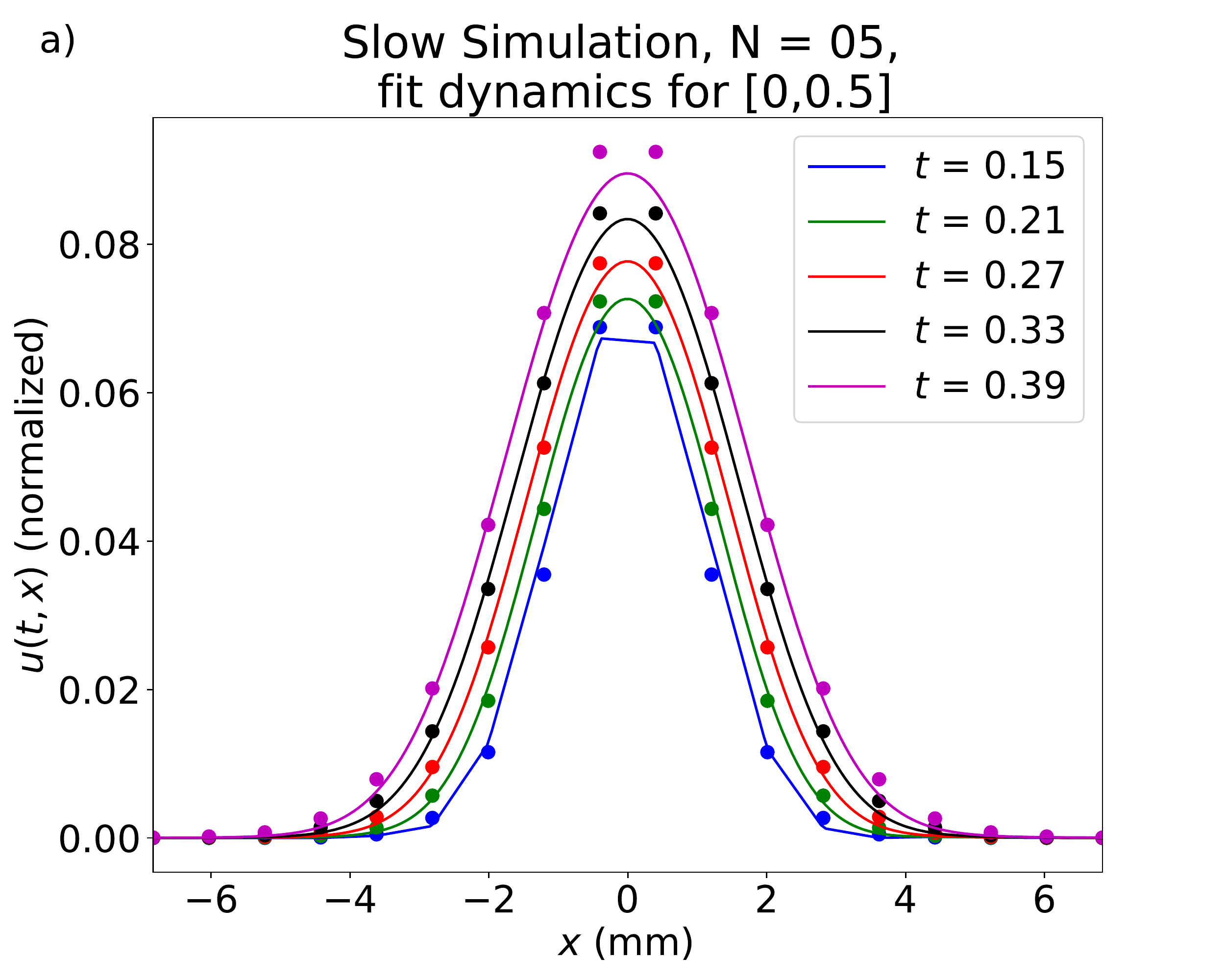}
    \includegraphics[width=0.45\textwidth]{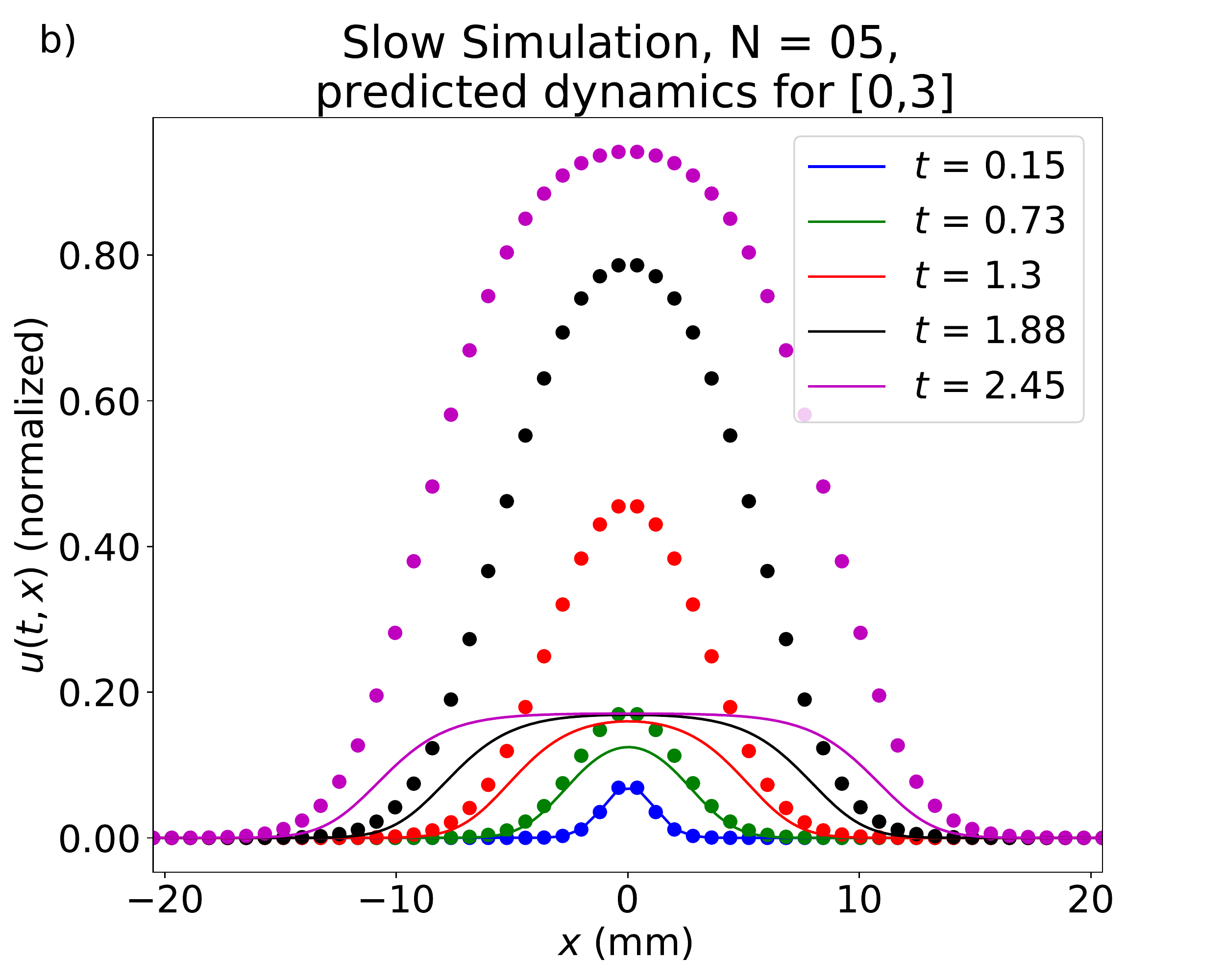}
    \includegraphics[width=0.45\textwidth]{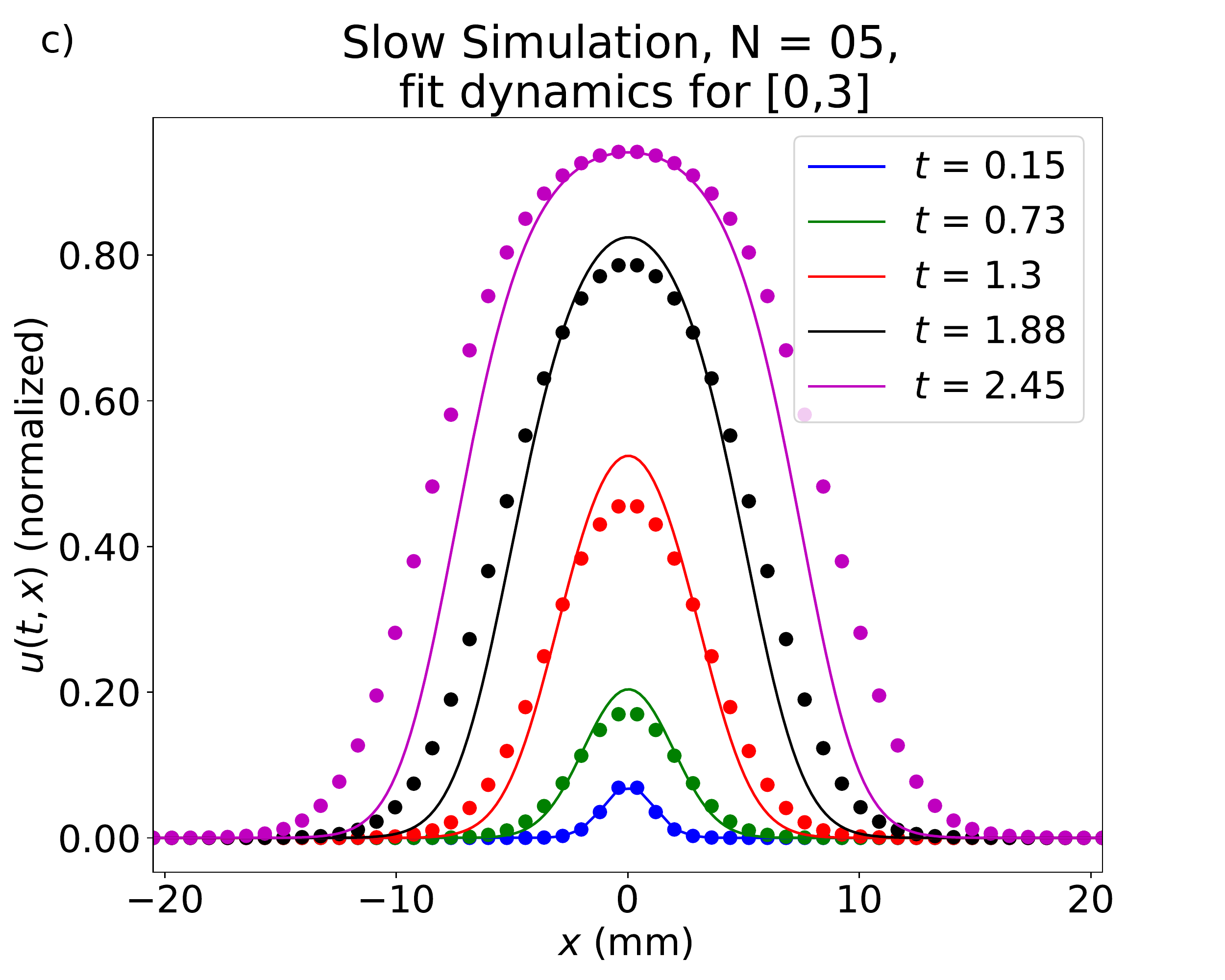}
    \includegraphics[width=0.45\textwidth]{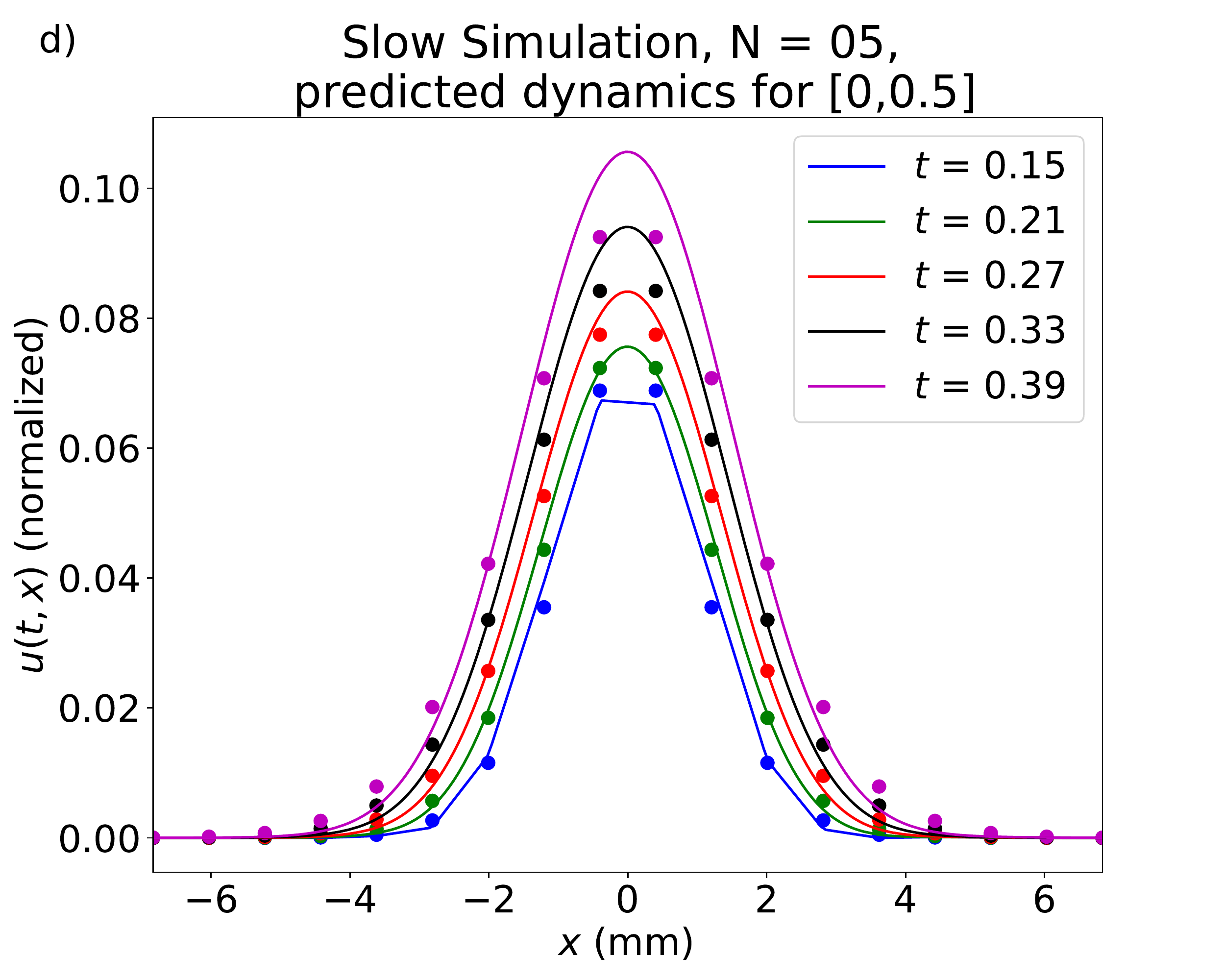}
    \caption{Fit and predicted dynamics for the slow simulation with $N=5$ time samples and $1\%$ noise. (a) The simulated learned Equation for the Slow simulation that was inferred from data sampled over the time interval [0,0.5]. (b) The model that was inferred over the time interval [0,0.5] is used to predict the dynamics over the time interval [0,3]. (c) The simulated learned Equation for the Slow simulation that was inferred from data sampled over the time interval [0,3]. (d) The model that was inferred over the time interval [0,3] is used to predict the dynamics over the time interval [0,0.5].
    }
    \label{fig:FKPP_generalization}
\end{figure}

\subsubsection{Learning equations from data with 1\% noise}\label{subsubsec:1perc}

All final inferred equations for 1\% noisy data are presented in Table \ref{tab:EQL_01}. Table cells are highlighted in green when the correct underlying equation form (i.e., the Fisher-KPP Equation) is inferred. We also tested the ability of the inferred equations with $N=5$ time samples to match the dynamics they were trained on and predict dynamics on a separate time interval (Figures \ref{fig:FKPP_generalization} and \ref{fig:FKPP_generalization_1}-\ref{fig:FKPP_generalization_3} in Appendix \ref{app:fit_predicted_dynamics}). 
\paragraph{The slow simulation on the short time scale.}  For noisy data sampled over the short time interval for the slow simulation, our equation learning methodology infers the Fisher-KPP Equation with $N=5$ and $10$ time samples. With $N=3$ points, the Fisher-KPP Equation is not inferred. In Figure \ref{fig:FKPP_generalization}a, we simulated the inferred equation for $N=5$ time samples over the short time scale and observe that this equation accurately matches the true underlying dynamics. When this same equation is simulated over the long time scale, it does not accurately describe the true underlying dynamics (Figure \ref{fig:FKPP_generalization}b). 
\paragraph{The slow simulation on the long time scale.} For noisy data sampled over the long time interval for the slow simulation, our equation learning methodology infers the Fisher-KPP Equation for all values of $N$ considered. In Figure \ref{fig:FKPP_generalization}c we simulate the inferred equation for $N=5$ time samples over the long time scale. The simulation matches the true dynamics well in many areas, although there is some disparity at $t=1.88$ and $2.45$ years for $|x|\ge 0.75$. When this same equation is simulated over the short time scale it accurately matches the true underlying dynamics (Figure \ref{fig:FKPP_generalization}d). 

\paragraph{The diffuse simulation on the short time scale.} For noisy data sampled over the short time interval for the diffuse simulation, our equation learning methodology does not infer the correct underlying equation for any of the chosen values of $N$. In Figure \ref{fig:FKPP_generalization_1} in Appendix \ref{app:fit_predicted_dynamics}, we simulated the inferred equation for $N=5$ time samples over the short time scale and observe that this equation does not match the true underlying dynamics. When this same equation is simulated over the long time scale, it does not accurately describe the true underlying dynamics. 

\paragraph{The diffuse simulation on the long time scale.} For noisy data sampled over the long time interval for the diffuse simulation, our equation learning methodology infers the Fisher-KPP Equation for all values of $N$ considered. In Figure \ref{fig:FKPP_generalization_1}, we simulate the inferred equation for $N=5$ time samples over the long time scale. The simulation matches the true dynamics qualitatively well in many areas. When this same equation is simulated over the short time scale, it accurately matches the true underlying dynamics. 

\paragraph{The fast simulation on the short time scale.} For noisy data sampled over the short time interval for the fast simulation, our equation learning methodology infers the Fisher-KPP Equation for $N=5$ and $10$ and does not infer the Fisher-KPP Equation for $N=3$. In Figure \ref{fig:FKPP_generalization_2} in Appendix \ref{app:fit_predicted_dynamics}, we simulated the inferred equation for $N=5$ time samples over the short time scale and observe that the inferred equation accurately describes the true underlying dynamics. When this same equation is simulated over the long time scale, it accurately matches the true underlying dynamics. 

\paragraph{The fast simulation on the long time scale.} For noisy data sampled over the long time interval for the fast simulation, our equation learning methodology does not infer the correct underlying equation for any of the chosen values of $N$. In Figure \ref{fig:FKPP_generalization_2} in Appendix \ref{app:fit_predicted_dynamics}, we simulated the inferred equation for $N=5$ time samples over the long time scale and observe that this equation does not accurately describe the true underlying dynamics. When this same equation is simulated over the short time scale, it does not accurately describe the true underlying dynamics. 

\paragraph{The nodular simulation on the short time scale.} For noisy data sampled over the short time interval for the nodular simulation, our equation learning methodology infers the Fisher-KPP Equation for all values of $N$ considered. In Figure \ref{fig:FKPP_generalization_3} in Appendix \ref{app:fit_predicted_dynamics}, we simulated the inferred equation for $N=5$ time samples over the short time scale and observe that the inferred equation accurately describes the true underlying dynamics. When this same equation is simulated over the long time scale, it accurately matches the true underlying dynamics.

\paragraph{The nodular simulation on the long time scale.} For noisy data sampled over the long time interval for the nodular simulation, our equation learning methodology does not infer the correct underlying equation for any of the chosen values of $N$. In Figure \ref{fig:FKPP_generalization_3} in Appendix \ref{app:fit_predicted_dynamics}, we simulated the inferred equation for $N=5$ time samples over the long time scale and observe that this equation does not accurately describe the true underlying dynamics. When this same equation is simulated over the short time scale, it does not accurately describe the true underlying dynamics. 

\subsubsection{Learning equations from data with 5\% noise}\label{subsubsec:5perc}

All final inferred equations for 5\% noisy data are presented in Table \ref{tab:EQL_05} in Appendix \ref{sec:learning_1d_long_supp}. Table cells are highlighted in green when the correct underlying equation form (i.e., the Fisher-KPP Equation) is inferred. We will also test the ability of the inferred equations with $N=10$ time samples to match the dynamics they were trained on and predict dynamics on a separate time interval (Figure \ref{fig:FKPP_generalization_4} in Appendix \ref{app:fit_predicted_dynamics}). 

\paragraph{The slow simulation on the short time interval.} For noisy data sampled over the short time interval for the slow simulation, our equation learning methodology does not infer the correct underlying equation for any values of $N$ considered. Simulating the inferred equation for $N=10$ time samples over the short time scale does not lead to an accurate description of the true underlying dynamics on the short time interval or prediction of the true dynamics on the long time interval.

\paragraph{The slow simulation on the long time interval.} Over the long time interval, our equation learning methodology does infer the Fisher-KPP Equation with $N=10$ time samples. Simulating the inferred equation for $N=10$ time samples over the long time scale accurately matches the true underlying dynamics on the long time interval and accurately predicts the true dynamics on the short time interval.

\paragraph{The diffuse simulation on the short time interval.} For noisy data sampled over the short time interval for the diffuse simulation, our equation learning methodology does not infer the correct underlying equation for any values of $N$ considered. Simulating the inferred equation for $N=10$ time samples over the short time scale does not lead to an accurate description of the true underlying dynamics on the short time interval or prediction of the true dynamics on the long time interval. 

\paragraph{The diffuse simulation on the long time interval.} Over the long time interval, our equation learning methodology does infer the Fisher-KPP Equation with $N=3$ time samples. Simulating the inferred equation for $N=10$ time samples over the long time scale accurately matches the true underlying dynamics on the long time interval and accurately predicts the true dynamics on the short time interval (Figure \ref{fig:FKPP_generalization_4} in Appendix \ref{app:fit_predicted_dynamics}).

\paragraph{The fast simulation on the short time interval.} For noisy data sampled over the short time interval for the fast simulation, our equation learning methodology infers the Fisher-KPP Equation with $N=10$ time samples. Simulating the inferred equation for $N=10$ time samples over the short time scale accurately matches the true underlying dynamics on the short time interval and accurately predicts the true dynamics on the long time interval (Figure \ref{fig:FKPP_generalization_4} in Appendix \ref{app:fit_predicted_dynamics}).

\paragraph{The fast simulation on the long time interval.} Over the long time interval, our equation learning methodology does not infer the correct underlying equation for any values of $N$ considered. Simulating the inferred equation for $N=10$ time samples over the short long scale does lead to an accurate description of the true underlying dynamics on the long time interval or prediction of the true dynamics on the short time interval.

\paragraph{The nodular simulation on the short time interval.} For noisy data sampled over the short time interval for the nodular simulation, our equation learning methodology infers the Fisher-KPP Equation with $N=3$ time samples. Simulating the inferred equation for $N=10$ time samples over the short time scale does not lead to an accurate description of the true underlying dynamics on the short time interval or prediction of the true dynamics on the long time interval.

\paragraph{The nodular simulation on the long time interval.} Over the long time interval, our equation learning methodology infers the Fisher-KPP Equation with $N=10$ time samples. Simulating the inferred equation for $N=10$ time samples over the long time scale accurately matches the true underlying dynamics on the long time interval and accurately predicts the true dynamics on the short time interval.

\subsubsection{Learning equations for intermediate $(D,r)$ values} \label{subsec:EQL_param_sweep}

Recall that the four considered simulations in this study correspond to values representing somewhat extreme examples of estimated values from a cohort of GBM patients \cite{wang_prognostic_2009}. In this section, we investigate the performance of our equation learning methodology for intermediate values of $D$ and $r$ from 1\% noisy data with $N=5$ time samples. Here we are interested in if we can infer the correct equation form, and score the accuracy of an inferred model form using the True Positive Ratio (TPR) given by:
\begin{equation}
    TPR = \dfrac{TP}{TP + FP + FN},
\end{equation}
where ``TP" stands for true positives (nonzero terms in the final inferred equation that are nonzero), ``FP" stands for false positives (nonzero terms in the final inferred equation that are zero), and ``FN" stands for false negatives (zero terms in the final inferred equation that are nonzero). Note that a score of TPR = 1 indicates that the correct underlying equation form has been recovered and a TPR score less than 1 indicates that the incorrect underlying equation form has been recovered.

\begin{figure}
    \centering
    \includegraphics[width=0.48\textwidth]{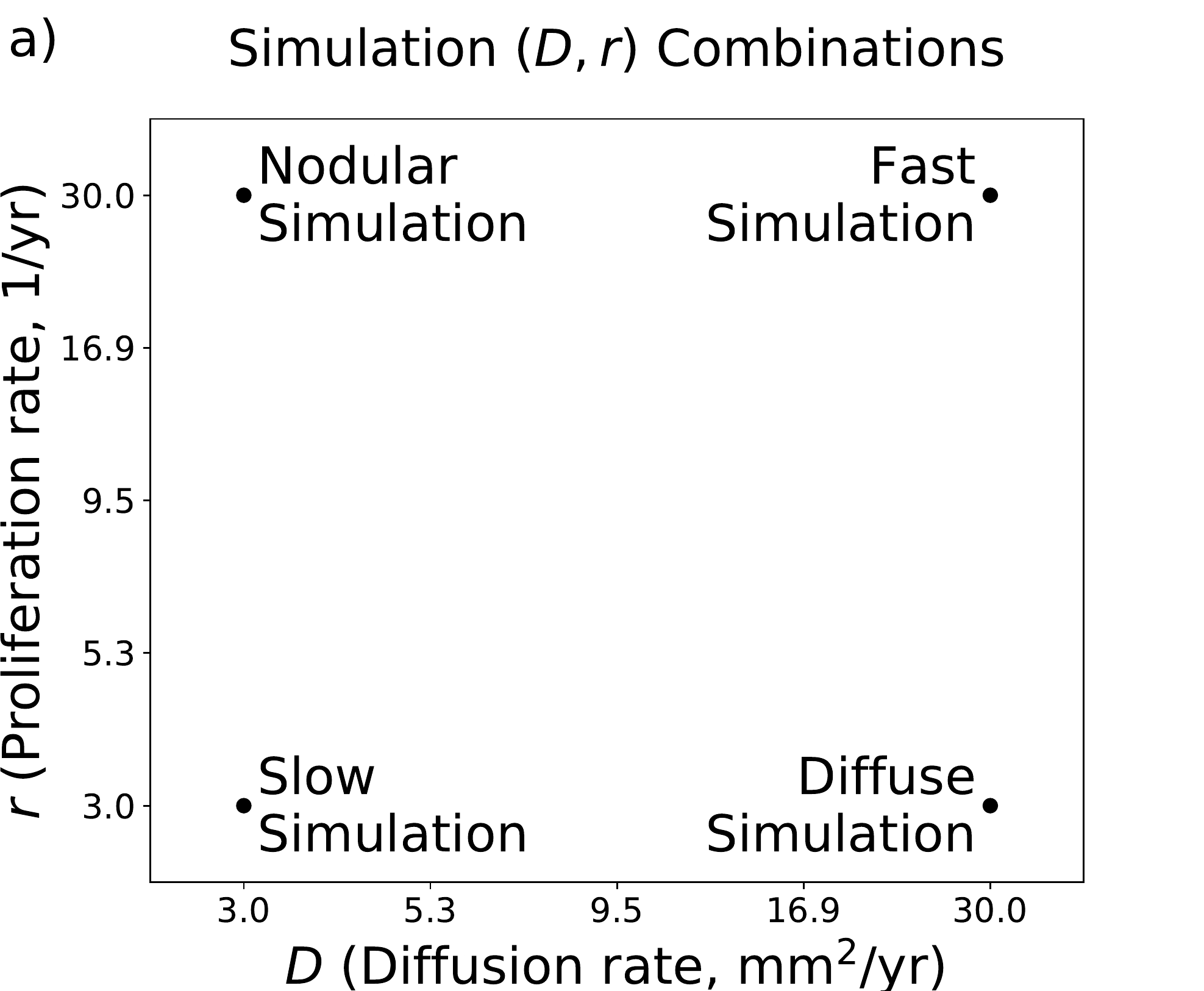}
    \includegraphics[width=0.48\textwidth]{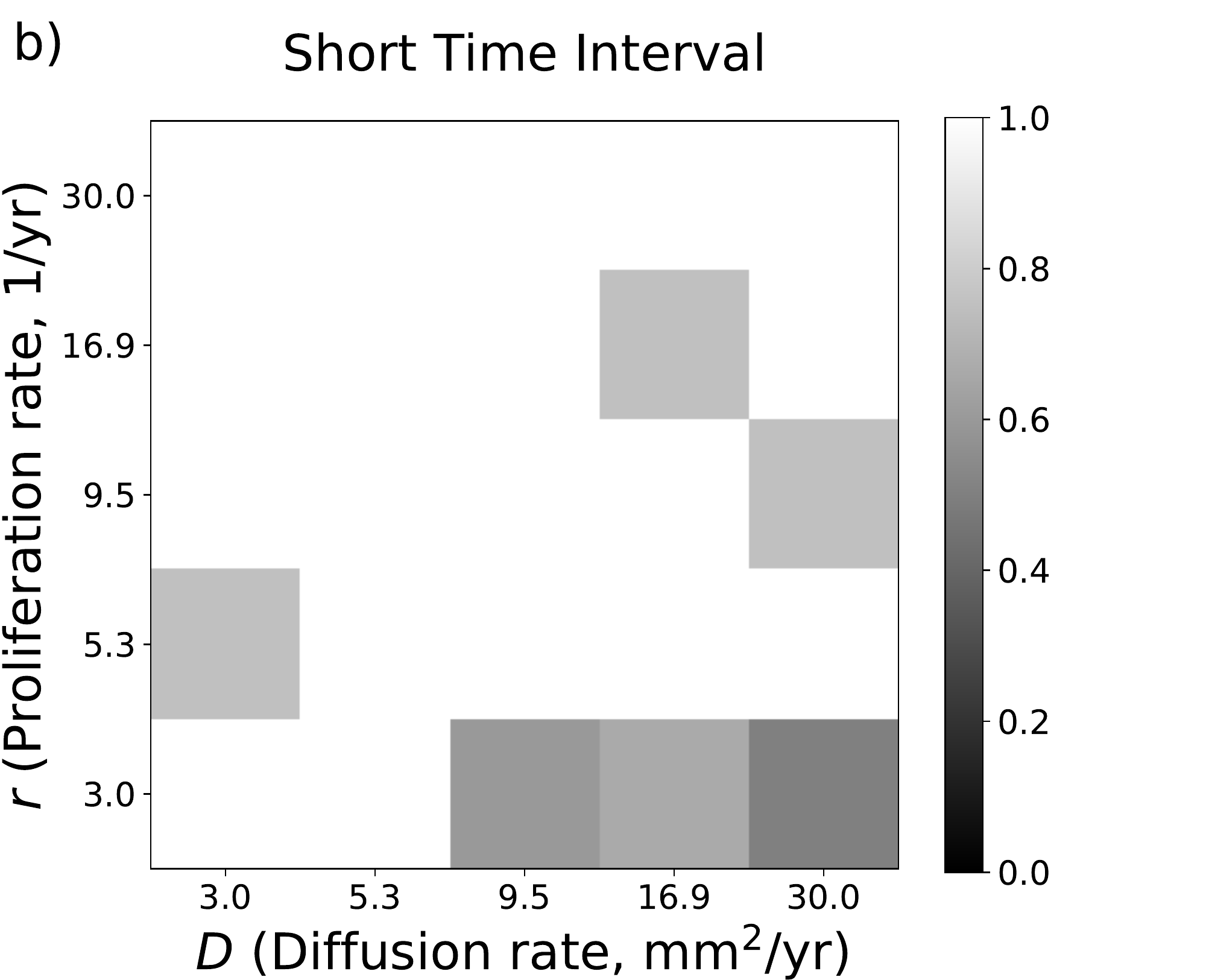}
    
    \includegraphics[width=0.48\textwidth]{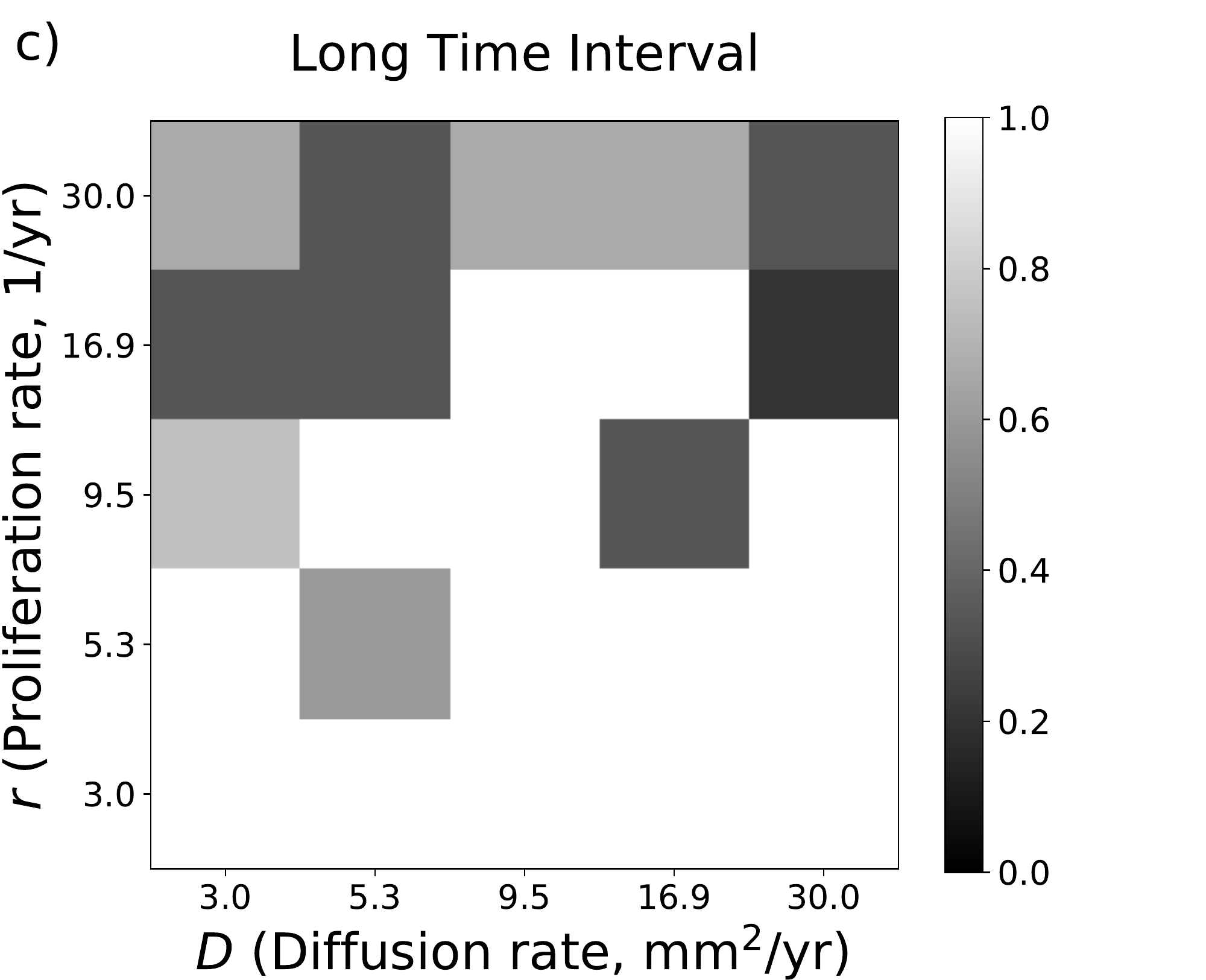}
    \caption{TPR scores over a range of $(D,r)$ values. (a) Depiction of where the Slow, Diffuse, Nodular, and Fast Simulations fall on these plots. (b) TPR scores for the 25 ($D,r$) combinations for data sampled on the short time interval (0-0.5 years). (c) TPR scores for the 25 ($D,r$) combinations for data sampled on the long time interval (0-3 years). }
    \label{fig:param_sweep}
\end{figure}

We inferred the equations underlying such data over 25 different $(D,r)$ combinations (Figure \ref{fig:param_sweep}). We let both $D$ and $r$ vary over five values from a log-scale between their lower and upper values from Table \ref{tab:variable_ranges}. Over the short time  interval, the correct underlying form is often inferred for larger values of $r$ ($r=5.3,9.5,16.9,30.0$ /year), but is not usually inferred for $r=3.0$/year. Over the long time interval, the correct underlying form is often inferred for smaller values of $r$ ($r=3.0,5.3,9.5$ /year), but is not usually inferred for $r=16.9$ or 30.0 /year.

\subsection{Parameter Estimation and Uncertainty Quantification} \label{sec:results_params_UQ}

We investigated parameter estimation accuracy and uncertainty quantification using the PDE-FIND algorithm when we know that the underlying form is the Fisher-KPP Equation. Results using data from the short time interval are presented in Section \ref{sec:PE_UQ_short} and the long time interval in Section \ref{sec:PE_UQ_long}.

\subsubsection{Parameter Estimation and Uncertainty Quantification over the short time interval}\label{sec:PE_UQ_short}

We investigated the performance of the PDE-FIND algorithm for parameter estimation and uncertainty quantification with 5\% noise and $N=5$ time samples on the short time interval (Table \ref{tab:UQ_short}) for the parameters $D$ and $r$. The fast simulation has the most accurate parameter estimation results on this time interval with error rates for $D$ and $r$ of 4.7\% and 0.5\%, respectively. 
The nodular simulation has error rates for $D$ and $r$ of 19.8\% and 4.7\%, respectively. The error rates for $r$ from the slow and diffuse simulations have error rates of 30.4\% and 21.8\%, respectively. PDE-FIND estimates $D$ with an error rate of 69.1\% for the slow simulation  and 25.1\% for the diffuse simulation. The normalized standard errors, defined as the standard error of the parameter estimates divided by their median value, for $r$ fall between 0.012-0.062 for all four simulations, and the more inaccurate estimates tend to exhibit higher normalized standard errors here. The normalized standard errors for $D$ are higher, as they fall between 0.084-.25 for all four simulations, demonstrating that this method exhibits higher variation in estimating $D$ than in estimating $r$. The fast simulation yields the smallest normalized standard error of 0.084 while the slow simulation yields the highest normalized standard error of 0.25.

\begin{table}
\centering
\begin{tabular}{|c|c|c|c|c|c|}
\hline
 & Parameter & True & Median & \% Error & Normalized SE \\ 
\hline
\multirow{2}*{\textbf{Slow Simulation}} & $D$ & 3 & 0.928 & 69.1\% & 0.25\\
 & $r$ & 3 & 2.089  & 30.4\% & 0.059\\
\hline
\multirow{2}*{\textbf{Fast Simulation}} & $D$ & 30 & 31.422 & 4.7\% & 0.084\\
 & $r$ & 30 & 29.84  & 0.5\% & 0.012\\
\hline
\multirow{2}*{\textbf{Diffuse Simulation}} & $D$ & 30 & 22.462 & 25.1\% & 0.132\\
 & $r$ & 3 & 2.345  & 21.8\% & 0.062\\
\hline
\multirow{2}*{\textbf{Nodular Simulation}} & $D$ & 3 & 3.595 & 19.8\% & 0.227\\
 & $r$ & 30 & 28.588  & 4.7\% & 0.027\\
\hline
\end{tabular}
\caption{1d parameter estimation and uncertainty quantification results for all simulations with 5\% noisy  data on a short time interval (0-0.5 years). \label{tab:UQ_short}}
\end{table}

\subsubsection{Parameter Estimation and Uncertainty Quantification over the long time interval}\label{sec:PE_UQ_long}

We investigated the performance of the PDE-FIND algorithm for parameter estimation and uncertainty quantification with 5\% noise and $N=5$ time samples on the long time interval (Table \ref{tab:UQ_long}) for the parameters $D$ and $r$. The smallest parameter estimate error rates are found for the diffuse simulation with error rates of 4.4 and 11.8\% for $D$ and $r$, respectively. PDE-FIND estimates $r$ with small error rates between 6.0-9.4\% error for the remaining simulations. The error rates for the remaining simulations for $D$ fall between 25.4-32.7\%. The diffuse simulation yields the lowest normalized standard error for $r$ of 0.012, while the fast simulation yields the highest normalized standard error of 0.047. The diffuse simulation yields the smallest normalized standard error for $D$ of 0.101 while the fast and nodular simulations yield the highest normalized standard errors for $D$ of 0.248.

\begin{table}
\centering
\begin{tabular}{|c|c|c|c|c|c|}
\hline
 & Parameter & True & Median & \% Error & Normalized SE \\ 
\hline
\multirow{2}*{\textbf{Slow Simulation}} & $D$ & 3 & 2.063 & 31.2\% & 0.168\\
 & $r$ & 3 & 2.786  & 7.1\% & 0.02\\
\hline
\multirow{2}*{\textbf{Fast Simulation}} & $D$ & 30 & 37.624 & 25.4\% & 0.248\\
 & $r$ & 30 & 28.189  & 6.0\% & 0.047\\
\hline
\multirow{2}*{\textbf{Diffuse Simulation}} & $D$ & 30 & 28.681 & 4.4\% & 0.101\\
 & $r$ & 3 & 2.646  & 11.8\% & 0.012\\
\hline
\multirow{2}*{\textbf{Nodular Simulation}} & $D$ & 3 & 2.02 & 32.7\% & 0.248\\
 & $r$ & 30 & 27.166  & 9.4\% & 0.023\\
\hline
\end{tabular}
\caption{1d parameter estimation and uncertainty quantification results for all simulations with 5\% noisy  data on a long time interval (0-3 years). \label{tab:UQ_long}}
\end{table}

\section{Discussion}\label{sec:discussion}

We investigated the performance of our equation learning methodology in equation inference, dynamics prediction, and parameter estimation from noisy data over a range of common challenges presented by biological data. These challenges include a wide variation in parameter values, sparse data sampling, and large amounts of noise. We used artificial data that has been generated from the Fisher-KPP Equation throughout this study due to the broad applicability of this model for many biological phenomena, including tumor progression and species invasion. The diffusion and proliferation values considered in this work correspond to the ranges of measured values from GBM patients \cite{wang_prognostic_2009}.

We observe in Table \ref{tab:EQL_01} that this methodology successfully recovers the correct underlying equation for the slow and diffuse simulations when data is observed on a long time interval of 0-3 years. The correct underlying form can be recovered with as few as three time samples on this time interval. Over the shorter time interval of 0-0.5 years, however, this methodology often either infers an incorrect equation form or infers the correct equation form with poor parameter estimates for these two simulations. With ten time samples from the slow simulation, for example, this methodology infers an equation with a carrying capacity that is ten times smaller than the true carrying capacity. Similarly, the inferred equation for the diffuse simulation will grow unbounded. These two simulations share a low proliferation rate of 3/year, suggesting that these simulations only exhibit their linear dynamics on the short interval and thus require the longer time interval for equation learning methods to accurately infer their nonlinear dynamics. These conclusions are supported by observing that the inferred equations from data that is sampled over the longer time interval can still accurately predict the system dynamics on the shorter time interval.

Alternatively, our equation learning methodology successfully recovers the correct underlying equations for the fast and nodular simulations over the shorter time interval of 0-0.5 years.   These two simulations share a high growth rate (30/year), suggesting that the systems' nonlinear dynamics are sufficiently sampled over the shorter time interval for successful equation recovery from equation learning methods. The inferred equations on this shorter time interval generalize well to predict the true underlying dynamics on the long time interval of 0-3 years. Our equation learning methodology does not infer the correct underlying equation for these two simulations when data is sampled over the long time interval. We observe in Figure \ref{fig:data} that these simulations appear to have converged to their traveling wave profiles between the first and second time samples of the long time interval. Traveling waves are a known problem for the PDE-FIND algorithm because multiple reaction-diffusion-advection equations will lead to traveling wave solutions. A previous study proposed simulating bimodal data to distinguish between the advection equation and the Korteweg-de Vries Equation \cite{rudy_data-driven_2017}. Here, we propose that sampling the dynamics before a traveling wave profile has been attained can lead to accurate equation inference for the Fisher-KPP Equation.




For low noise ($\sigma=1\%$) data, the final inferred equation form appears robust to the number of time samples. When data is observed over the correct time interval, then the final equation form typically does not change between three and ten time samples (Table \ref{tab:EQL_01}). An exception is the fast simulation, which needed five or ten time samples to recover the correct underlying equation. With a larger amount of noise ($\sigma=5\%$), the final inferred equation appears more sensitive to the the number of observed time samples. For example, the slow simulation required five or ten time samples to infer the correct underlying equation and the fast simulation required ten time samples to infer the correct underlying equation. Interestingly, our equation learning methodology inferred the correct equation for the diffuse and nodular simulations with three time samples with 5\% noise but inferred incorrect equation forms for these two simulations with five or ten time samples. We note, however, that the extra terms for the diffuse simulation here correspond to backwards negative nonlinear diffusion and advection which a user may manually neglect if found in practice. Furthermore, we could have had more success recovering the correct equation form by tuning hyperparameters (such as the pruning percentage) here, but we instead focus on a flexible technique that provides users with interpretable equations for users to then alter if needed.

The four simulations considered throughout this work correspond to the outer ranges of diffusion and proliferation observed in GBM patients. We further demonstrated in Figure \ref{fig:param_sweep} that our equation learning methodology is successful in recovering the correct equation learning methodology for many intermediate $(D,r)$ values. Typically, our methodology has more success in accurate equation inference for larger values of $r$ ($r\ge5.3$/year) on the short time interval and for smaller values of $r$ ($r\le9.5$/year) on the longer time interval for 1\% noisy data.

Equation learning is promising technique for parameter estimation and uncertainty quantification due to having a low computational expense. For example, solving Equation \eqref{eq:PDE-FIND} for $\xi$ does not require solving a differential equation, whereas typical parameter estimation routines require numerous differential equation model simulations \cite{banks2014modeling}. We observe that parameter estimation appears more accurate for the slow and diffuse simulations over the long time interval than for the short time interval. Similarly, parameters are more accurate for the fast and nodular simulations over the short time interval than the long time interval. Such methods could be used to obtain computationally inexpensive initial parameter estimates for frequentist parameter estimation approaches approaches or for prior distribution specification for Bayesian inference routines \cite{hawkins-daarud_quantifying_2019}. Parameter uncertainty has been proposed previously as a measure to infer the top equation form from several plausible candidates \cite{zhang2018robust}. In support of this proposal, we observe in our work that smaller amounts of parameter uncertainty result when data is sampled over an appropriate time interval.

\section{Conclusion}\label{sec:conclusion}


Methods from equation learning have the potential to become invaluable tools for researchers in developing data-driven models, and in turn understanding the mechanics underlying a biological process, estimating parameters, and predicting previously-unknown system dynamics. Before such methods become commonplace in the mathematical biology field, however, it is crucial to thoroughly investigate the performance of such methods in the presence of common challenges that biological data presents. We have scrutinized the performance of a state-of-the-art equation learning methodology in the presence of such challenges in this work. This equation learning methodology is composed of a data denoising step with an ANN, equation learning through the PDE-FIND algorithm \cite{lagergren_learning_2020,rudy_data-driven_2017}, and a final step comprised of model selection and post-processing to ensure the inferred equation is simple and biologically interpretable.

The biological data challenges considered in this work include sparse sampling of data, a small number of time samples, parameter heterogeneity, and large noise levels. Our equation learning method can recover the correct equation from data with a small number of time samples when the data is sampled over an appropriate time interval. Our methodology can also reliably predict previously-unobserved dynamics when trained over an appropriate time interval. When this methodology is not trained over an appropriate time interval, however, the inferred equation and predicted dynamics are not reliable. Determining when to sample the data for accurate inference in this work crucially depends on the intrinsic growth rate of populations: fast-growing populations require data sampling on a short time interval and slow-growing populations require sampling on a long time interval for accurate inference. When sampled over the correct time interval, datasets exhibited a combination of both initial linear model dynamics and long-term nonlinear model dynamics. Such results suggest that an informative time interval for equation learning methods should include both of these phases for accurate inference. Noisier data requires more time samples for accurate recovery: we observed in this work that three time samples were often sufficient for accurate inference of 1\% noisy data, but ten time samples were required for such inference on 5\% noisy data. Deciphering when equation learning methodologies are reliable in the presence of practical data challenges is important for biological research to ensure these methods are not used to make incorrect inference for a system under study. The challenges addressed in this study are prevalent in biological research, where expensive or challenging data collection may limit us to sparse datasets \cite{baldock_patient-specific_2014} or measurement is often corrupted by noise \cite{perretti_model-free_2013}.

\begin{figure}
    \centering
    \includegraphics[width=0.48\textwidth]{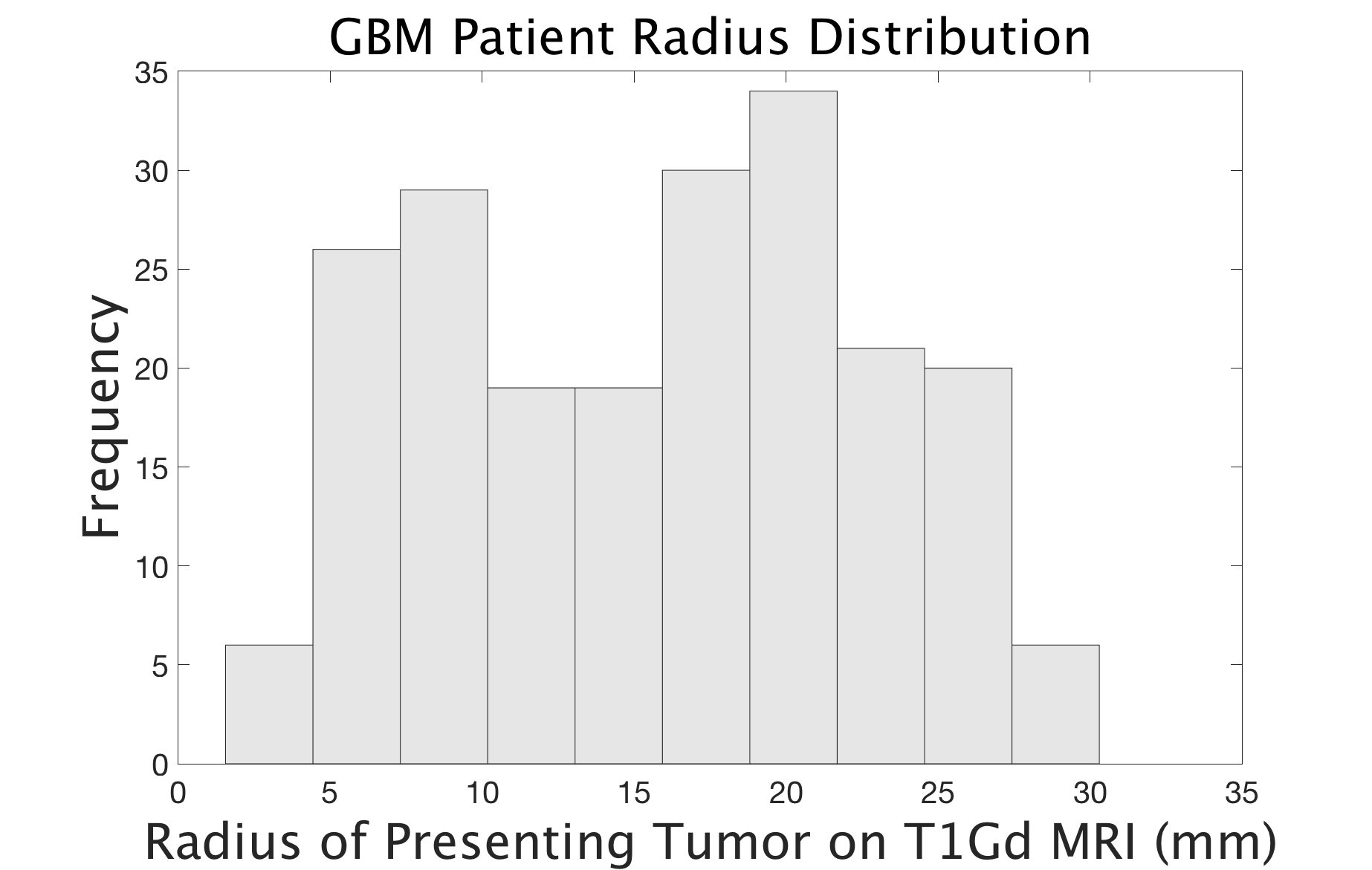}
    \caption{Histogram of measured GBM patient radii when first presenting in the clinic obtained from T1Gd imaging.}
    \label{fig:histograms_2}
\end{figure}

The values of diffusion and intrinsic growth considered in this work correspond to the ranges of these values that have been measured from GBM patients' MR images \cite{wang_prognostic_2009}. In Figures \ref{fig:histograms} and \ref{fig:histograms_2}, we depict histograms of measured $D/r$ ratios, T1Gd velocities, and tumor radii from GBM patients before surgery. If one is interested in inferring the dynamics from patient data to inform treatment decisions for improved prognosis or predicted dynamics, then patient-estimated $D/r$ tumor velocity measurements can be combined with the results from this study to determine how reliable the inferred equation may be. A patient's tumor could be matched to one of the four simulations considered in this work based on their measured $D/r$ and velocity estimates (for example, if the tumor has a high $D/r$ value  and low velocity, then this tumor may correspond to the slow simulation). If patient data has been observed on an appropriate time interval for inference (a longer time interval for slow and diffuse simulations, a shorter time interval for fast and nodular simulations), then one may have a high degree of confidence in the inferred equation and dynamics. If patient data has not been observed on an appropriate time interval for inference, then the inferred equation and dynamics might not be sufficiently supported by the data. We also observed in this work that if the population dynamics are primarily observed after the population has reached confluence, then the inferred dynamics may not be reliable. The tumor radius histogram in Figure \ref{fig:histograms_2} can be used to further determine whether or not observed patient data has already reached its carrying capacity and is suitable for inference from equation learning methods.

We focused on inference of the Fisher-KPP Equation in this work due to its wide use in the biological literature \cite{hastings_spatial_2005,hawkins-daarud_quantifying_2019,nardini_investigation_2018}. Previous equation learning studies have also successfully inferred other common models in biology, including transport equations \cite{lagergren_learning_2020}, pattern-forming reaction-diffusion equations \cite{rudy_data-driven_2017}, and compartmental epidemiological models \cite{mangan2017model}. We expect that the results, methodology, and open-source code presented in this work will enable other researchers to assess whether equation learning methods can be applied to a wide array biological data. 

This work proposes many areas for future research for mathematicians and data scientists working on computational methods and modeling for biological data. We observed that changing the parameters of a mathematical model influences the spatiotemporal domain over which the PDE-FIND algorithm can successfully recover the correct underlying equation. Future research should aim to infer what an informative domain may be for general ODE and PDE systems as well as methods to determine if datasets contain high or low information content for equation learning. As an example, only sampling near an equilibrium solution will likely lead to a learned model of $u_t=0$, whereas only sampling far away from stable equilibria may neglect nonlinear dynamics in the final learned model. It will be interesting to further investigate what combinations of transient and long-term dynamics are sufficient for accurate inference from biological systems data, which may be difficult or expensive to collect. Improvements in data denoising and equation learning methods will further inform this work. A recent neural network architecture, termed a ``Physics-informed neural network", simultaneously denoises data and estimates the parameters under the assumption that the model is known \cite{raissi_hidden_2018}. The simultaneous denoising of data and equation inference will likely improve both methods and is thus an important area for future research.

\appendix

\section{Simulating a learned Equation} \label{sec:simulating_learned}

To simulate the inferred equation represented by the sparse vector $\hat{\xi}$, we begin by removing all zero terms from $\hat{\xi}$ as well as the corresponding terms from $\Theta$. We can now define our inferred dynamical systems model as 
\begin{equation}
    u_t = \sum_i \xi_i \Theta_i.
\end{equation}
We use the method of lines approach to simulate this equation, in which we discretize the right hand side in space and then integrate along the $t$ dimension. The Scipy integration subpackage (version 1.4.1) is used to integrate this equation over time using an explicit fourth order Runge-Kutte Method. We ensure that the simulation is stable by enforcing the CFL Condition for an advection equation with speed $2\sqrt{Dr}$ is satisfied, \emph{e.g.}, $2\sqrt{Dr}\Delta t \le \Delta x$. Some inferred equations may not be well-posed, \emph{e.g.}, $u_t=-u_{xx}$. If the time integration fails at any point, we manually set the model output to $10^6$ everywhere to ensure this model is not selected as a final inferred model.

For the final inferred columns of $\Theta=[\Theta_1 , \Theta_2 , \dots , \Theta_n]$, we define nonlinear stencils, $A_{\Theta_i}$ such that $A_{\Theta_i}u\approx \Theta_n$. As an example, we an upwind stencil \cite{leveque_finite_2007} for first order derivative terms, such as $A_{u_x}$, so that $A_{u_x}u\approx u_x$. We use a central difference stencil for $A_{u_{xx}}$. For multiplicative terms, we define the stencil for $A_{uu_x}$ as $A_{uu_x}v=u\odot (A_{u_x}v),$ where $\odot$ denotes element-wise multiplication so that $A_{uu_x}u \approx uu_x$. Similarly, we set $A_{u_xu_{xx}}=A_{u_x}A_{u_{xx}}$, etc. 


\section{Learning the 1d Fisher-KPP Equation with 5\% noisy data}\label{sec:learning_1d_long_supp}

In Table \ref{tab:EQL_05}, we present the inferred equations for all 1d data sets considered with $\sigma = 0.05$.


\begin{table}
\centering
\begin{tabular}{|c|c|c|c|}
\hline
\multicolumn{2}{|c|}{\textbf{Slow Simulation}} & \multicolumn{2}{c|}{$\bm{u_t = 3u_{xx} + 3u-3u^2}$} \\
\hline
$\sigma$ & \ $N$ & Learned Equation (0-0.5 years) & Learned Equation (0-3 years) \\ 
\hline
     05 & 03 & $u_t = -101.718u^2 + 9.941u$ & $u_t = 3.3u_{xx}-2.26u^2 + 2.501u-0.6uu_{x}$ \\
\hline
     05 & 05 & $u_t = 1.637u$ & $u_t = 1.9u_{xx} + 2.788u-2.8u^2$ \cellcolor [HTML]{77dd77} \\
\hline
     05 & 10 & $u_t = -35.4u^2 + 4.936u-15.2uu_{x}$ & $u_t = 2.2u_{xx}-2.913u^2 + 3.002u$ \cellcolor [HTML]{77dd77} \\
\hline
\multicolumn{4}{|c|}{ } \\
\hline
\multicolumn{2}{|c|}{\textbf{Diffuse Simulation}} & \multicolumn{2}{c|}{$\bm{u_t = 30u_{xx} + 3u-3u^2}$} \\
\hline
$\sigma$ & \ $N$ & Learned Equation (0-0.5 years) & Learned Equation (0-3 years) \\ 
\hline
     05 & 03 & $u_t = 2349.5u_{x}^2$ & $u_t = 24.2u_{xx}-4.034u^2 + 3.071u$ \cellcolor [HTML]{77dd77} \\
\hline
     \multirow{2}{*}{05} & \multirow{2}{*}{05} & \multirow{2}{*}{$u_t = -422.687u^2 + 9.883u$} & $u_t = 29.2u_{xx}-3.647u^2$ \\
      & & & $+ 3.254u-40.1uu_{xx}-98.6u_{x}^2$ \\
\hline
     \multirow{2}{*}{05} & \multirow{2}{*}{10} & \multirow{2}{*}{$u_t = 3551.9u_{x}^2$} & $u_t = -0.54u_{x} + 28.4u_{xx}-3.181u^2$ \\
      & & & $+ 2.996u + 2.9u^2u_{x}-36.5uu_{xx} + 22.2u_{x}^2$ \\
\hline
\multicolumn{4}{|c|}{ } \\
\hline
\multicolumn{2}{|c|}{\textbf{Fast Simulation}} & \multicolumn{2}{c|}{$\bm{u_t = 30u_{xx} + 30u-30u^2}$} \\
\hline
$\sigma$ & \ $N$ & Learned Equation (0-0.5 years) & Learned Equation (0-3 years) \\ 
\hline
     05 & 03 & $u_t = -30.113u^2 + 29.593u$ & $u_t = -155.4u^2u_{xx} + 644.4u_{x}^2$ \\
\hline
     \multirow{2}{*}{05} & \multirow{2}{*}{05} & $u_t = 40.1u_{xx}-28.118u^2 + 28.84u$ & \multirow{2}{*}{$u_t = 60.8u_{xx}-220.3u^2u_{xx}$} \\
      & & $+ 5.86u^2u_{x}$ & \\
\hline
     05 & 10 & $u_t = 23.5u_{xx}-29.686u^2 + 29.83u$ \cellcolor [HTML]{77dd77} & $u_t = -26.923u^2 + 26.964u$ \\
\hline
\multicolumn{4}{|c|}{ } \\
\hline
\multicolumn{2}{|c|}{\textbf{Nodular Simulation}} & \multicolumn{2}{c|}{$\bm{u_t = 3u_{xx} + 30u-30u^2}$} \\
\hline
$\sigma$ & \ $N$ & Learned Equation (0-0.5 years) & Learned Equation (0-3 years) \\ 
\hline
     05 & 03 & $u_t = 3.2u_{xx}-21.375u^2 + 22.857u$ \cellcolor [HTML]{77dd77} & $u_t = -34.4u^2u_{xx} + 73.3u_{x}^2$ \\
\hline
     05 & 05 & $u_t = 6.7u_{xx}$ & $u_t = -26.496u^2 + 26.634u$ \\
\hline
     05 & 10 & $u_t = 7.4u_{xx}$ & $u_t = 2.9u_{xx}-24.297u^2 + 24.427u$ \cellcolor [HTML]{77dd77} \\
\hline
\end{tabular}
\caption{Learned 1d Equations from our equation learning methodology for all simulations with 5\% noisy  data. Correctly-inferred equation forms are shaded in green. \label{tab:EQL_05}}
\end{table}

\section{Fit and Predicted Dynamics} \label{app:fit_predicted_dynamics}

\begin{figure}
    \centering
    \includegraphics[width=0.45\textwidth]{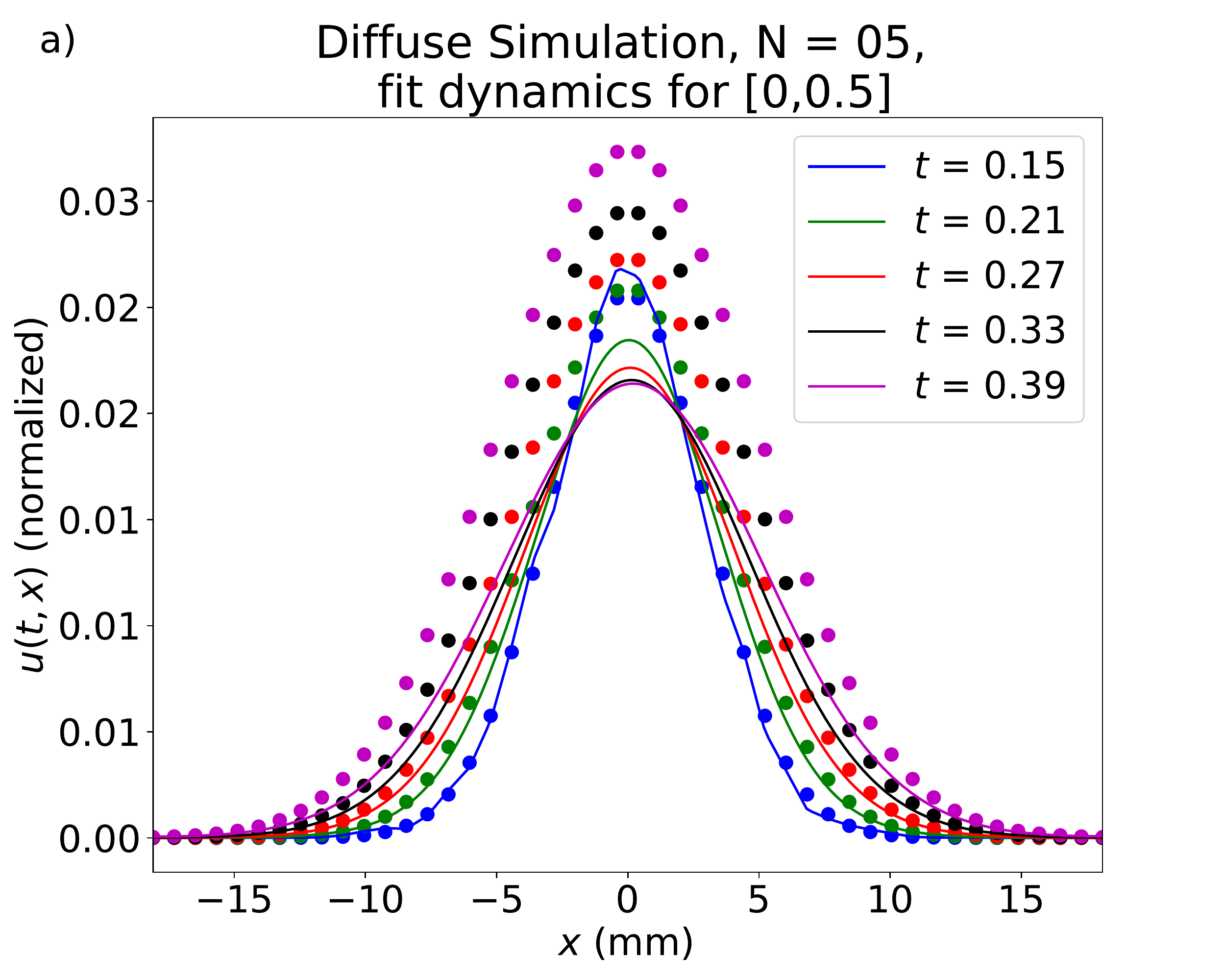}
    \includegraphics[width=0.45\textwidth]{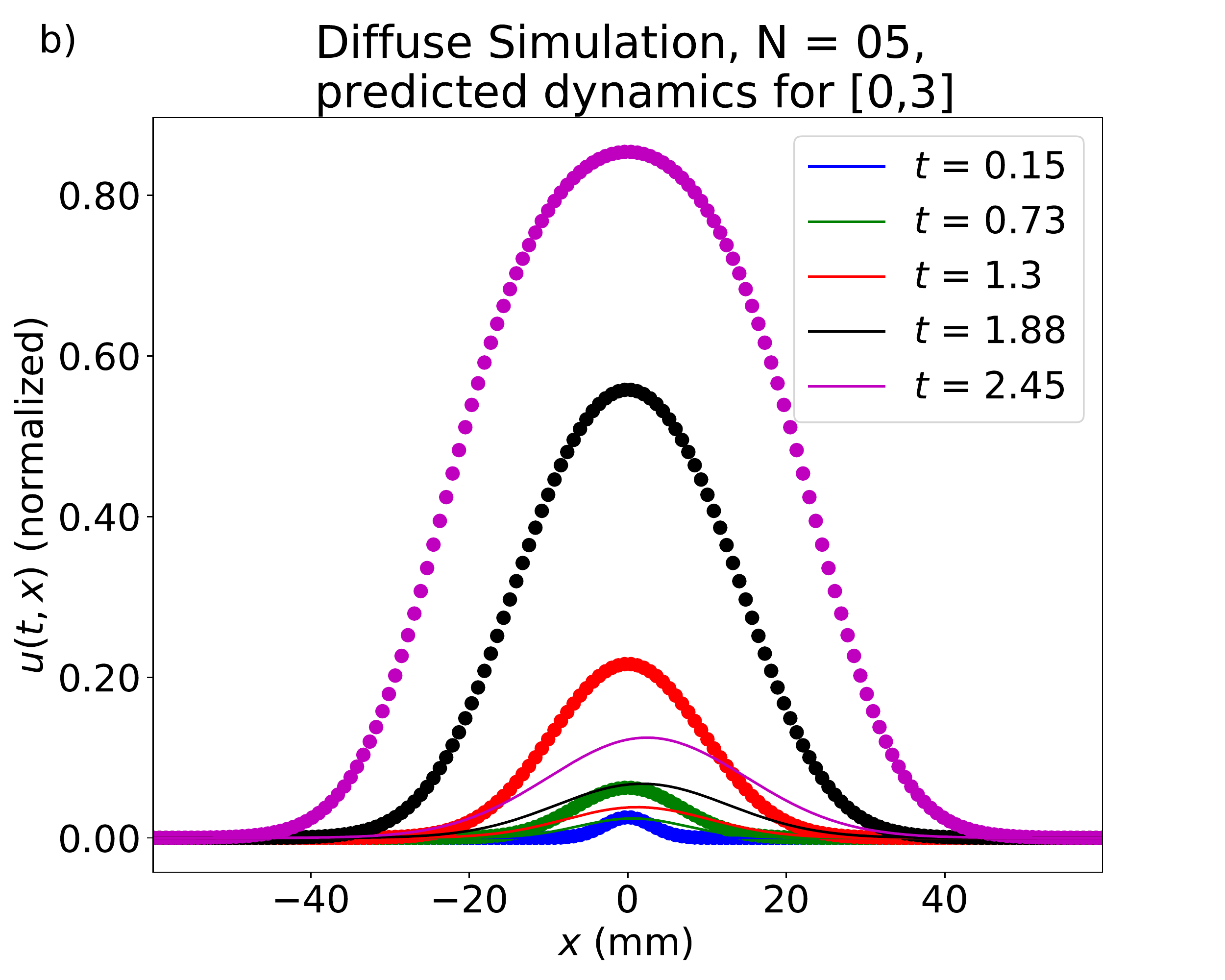}
    \includegraphics[width=0.45\textwidth]{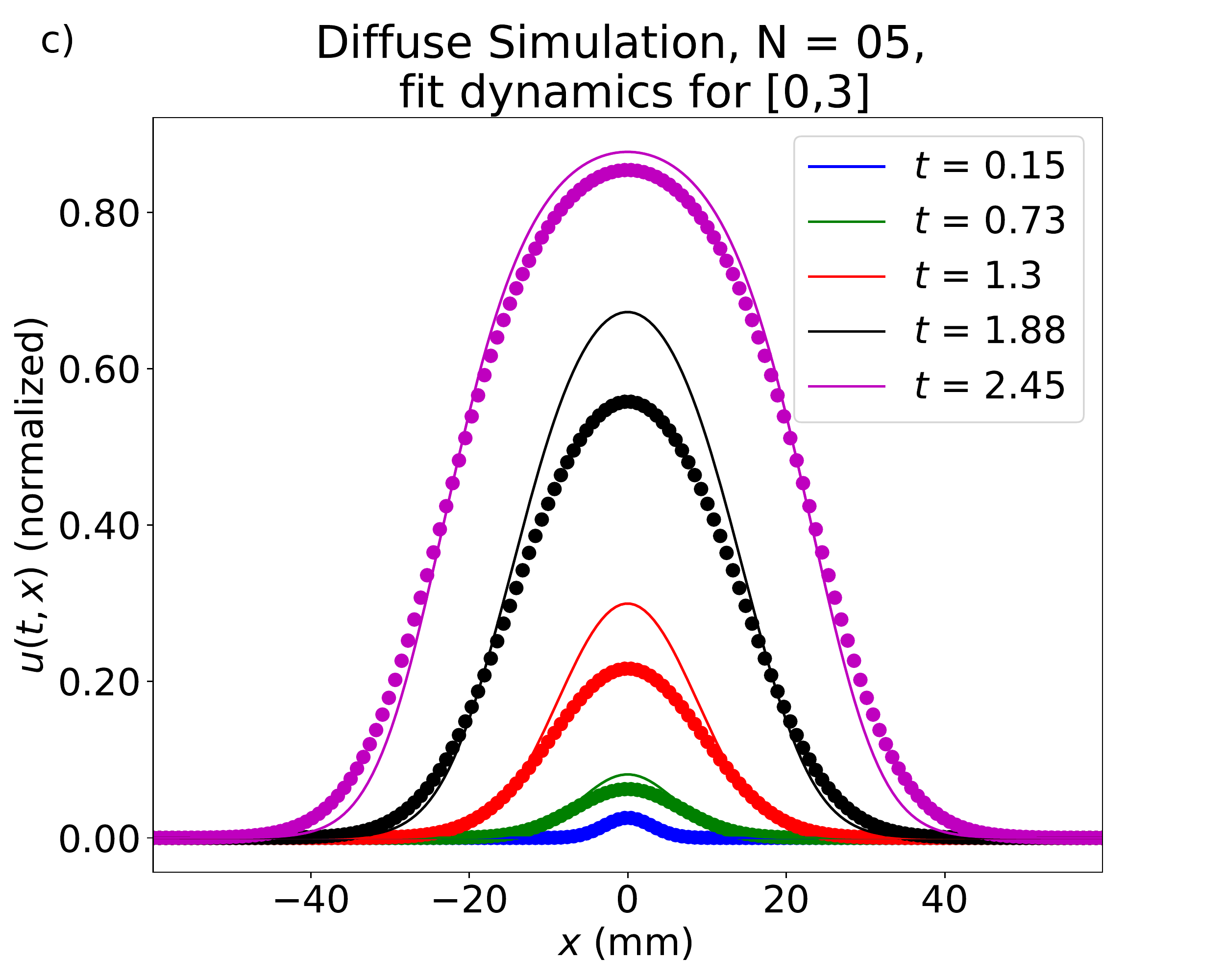}
    \includegraphics[width=0.45\textwidth]{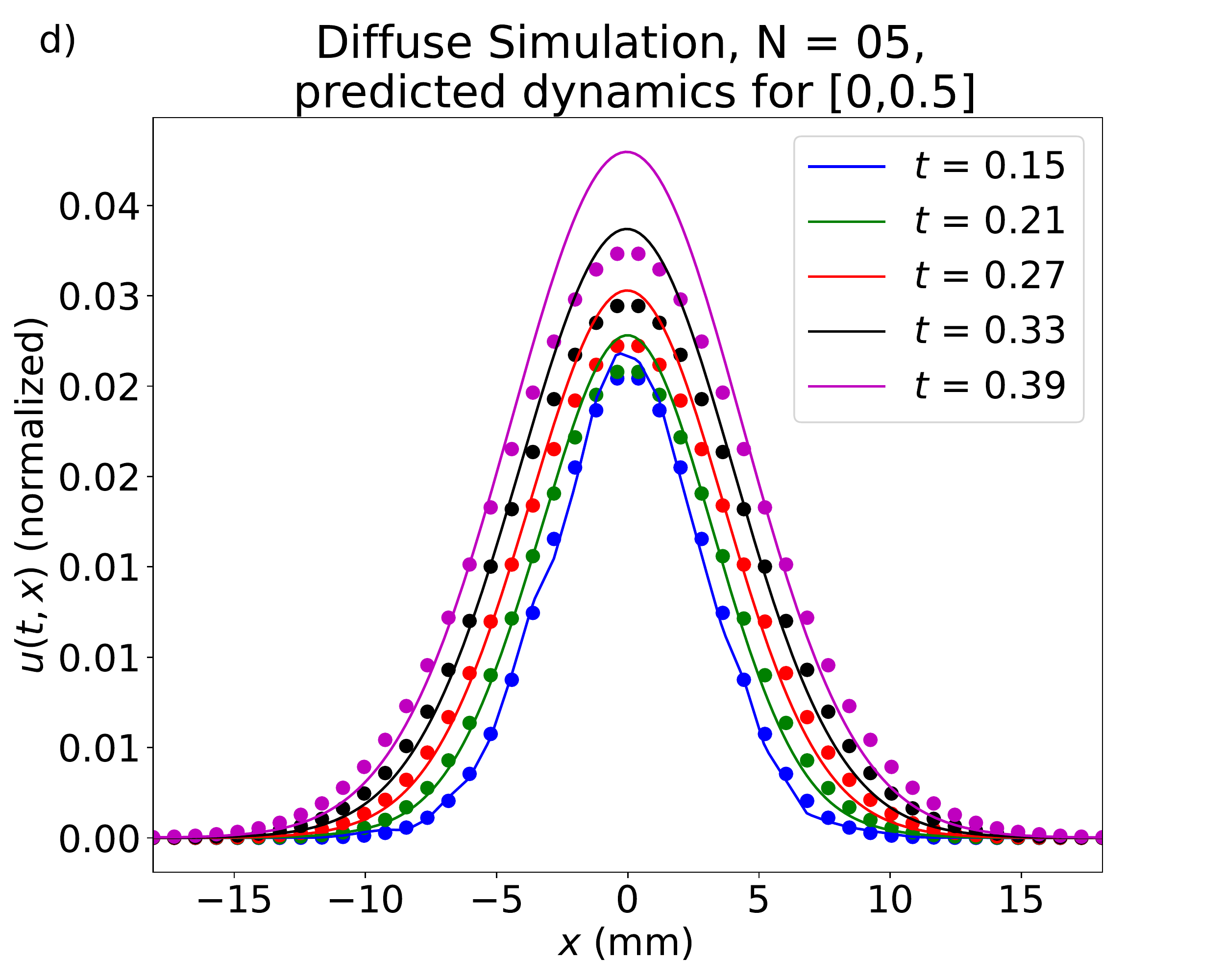}
    \caption{Fit and predicted dynamics for the fast  with $N=5$ time samples and $1\%$ noise. (a) The simulated learned Equation for the fast  that was inferred from data sampled over the time interval [0,0.5]. (b) The model that was inferred over the time interval [0,0.5] is used to predict the dynamics over the time interval [0,3]. (c) The simulated learned Equation for the fast  that was inferred from data sampled over the time interval [0,3]. (d) The model that was inferred over the time interval [0,3] is used to predict the dynamics over the time interval [0,0.5]. Simulated models are shown in solid lines and the true underlying dynamics are shown by dots.
    }
    \label{fig:FKPP_generalization_1}
\end{figure}

\begin{figure}
    \centering
    \includegraphics[width=0.45\textwidth]{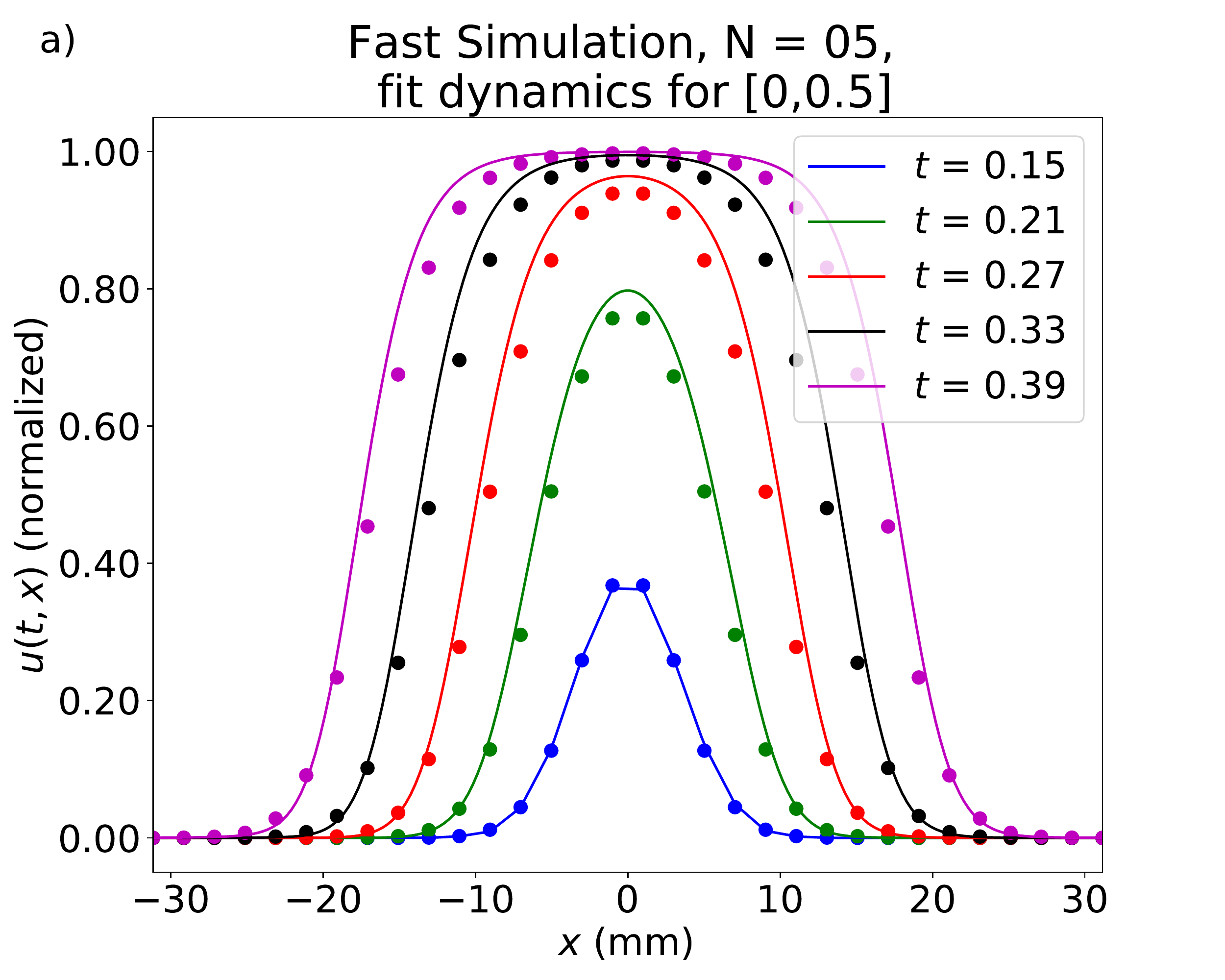}
    \includegraphics[width=0.45\textwidth]{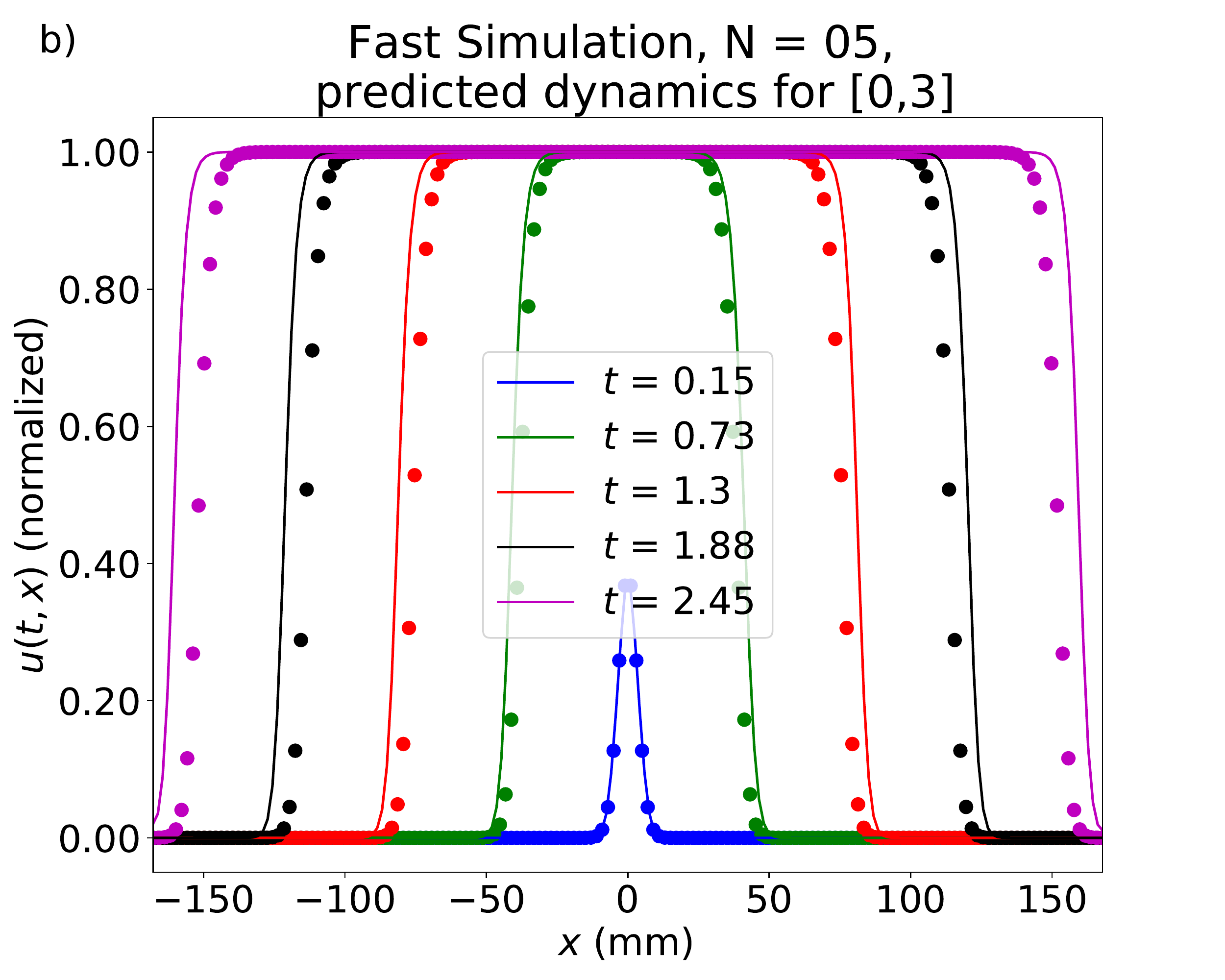}
    \includegraphics[width=0.45\textwidth]{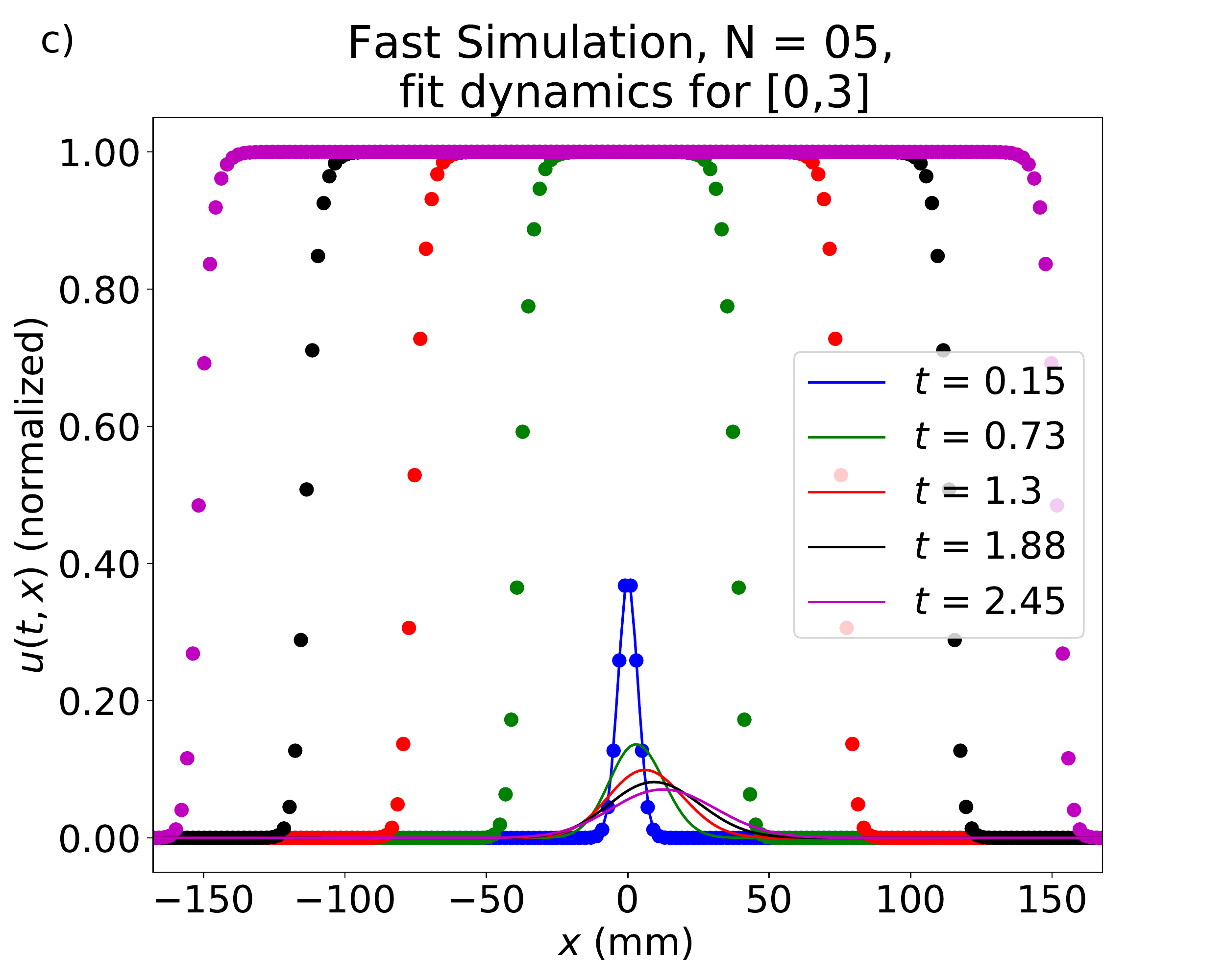}
    \includegraphics[width=0.45\textwidth]{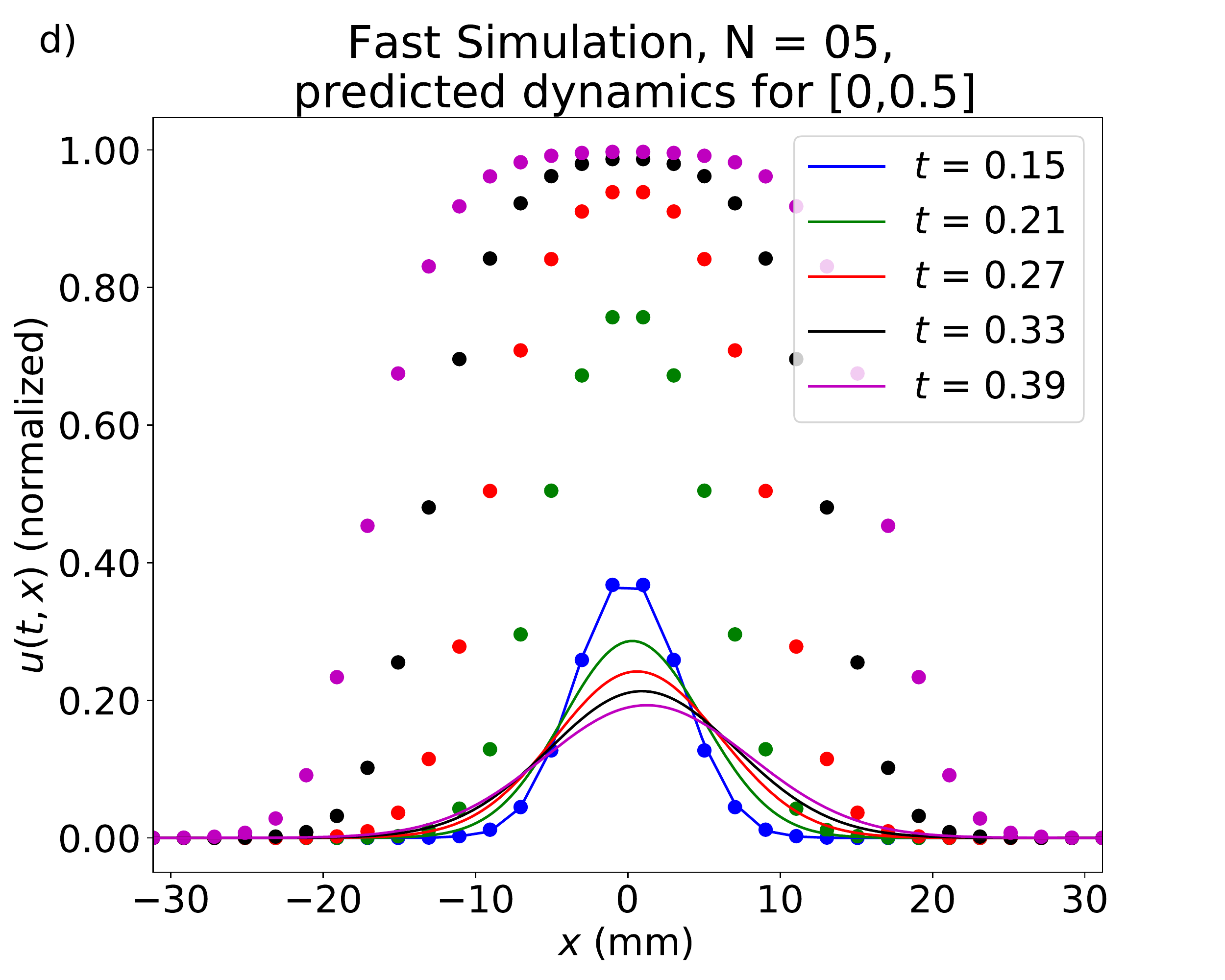}
    \caption{Fit and predicted dynamics for the diffuse  with $N=5$ time samples and $1\%$ noise. (a) The simulated learned Equation for the diffuse  that was inferred from data sampled over the time interval [0,0.5]. (b) The model that was inferred over the time interval [0,0.5] is used to predict the dynamics over the time interval [0,3]. (c) The simulated learned Equation for the diffuse  that was inferred from data sampled over the time interval [0,3]. (d) The model that was inferred over the time interval [0,3] is used to predict the dynamics over the time interval [0,0.5].
    }
    \label{fig:FKPP_generalization_2}
\end{figure}

\begin{figure}
    \centering
    \includegraphics[width=0.45\textwidth]{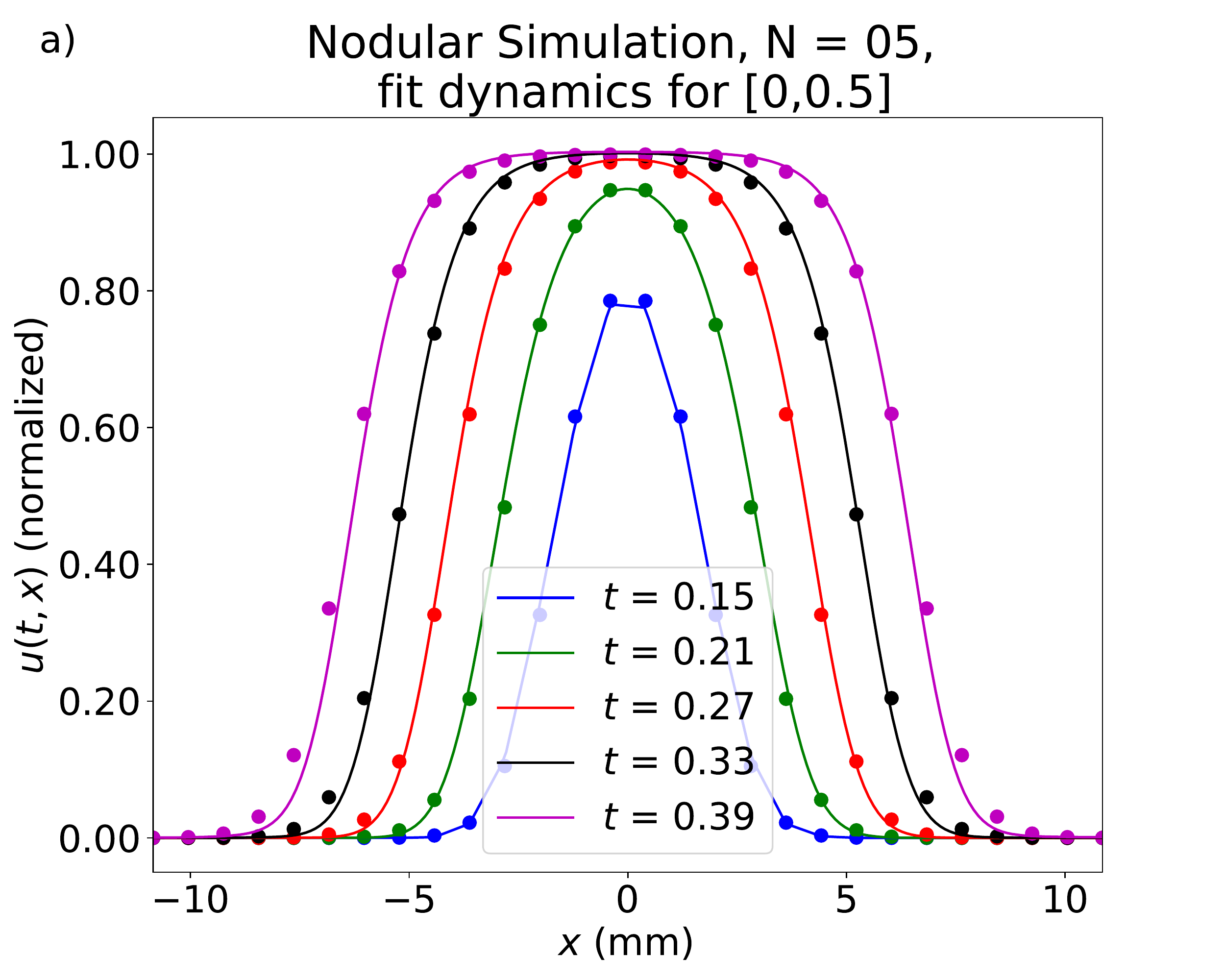}
    \includegraphics[width=0.45\textwidth]{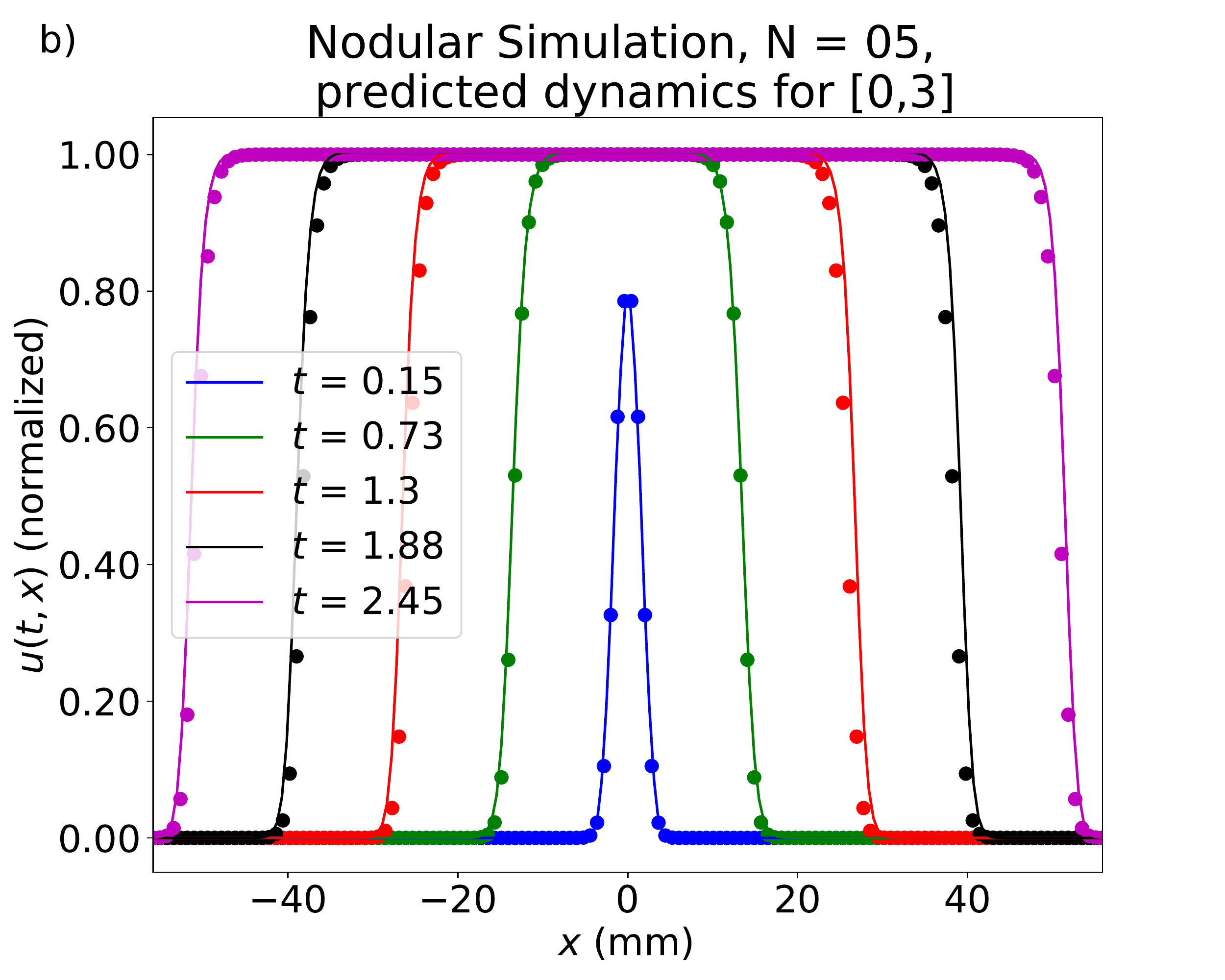}
    \includegraphics[width=0.45\textwidth]{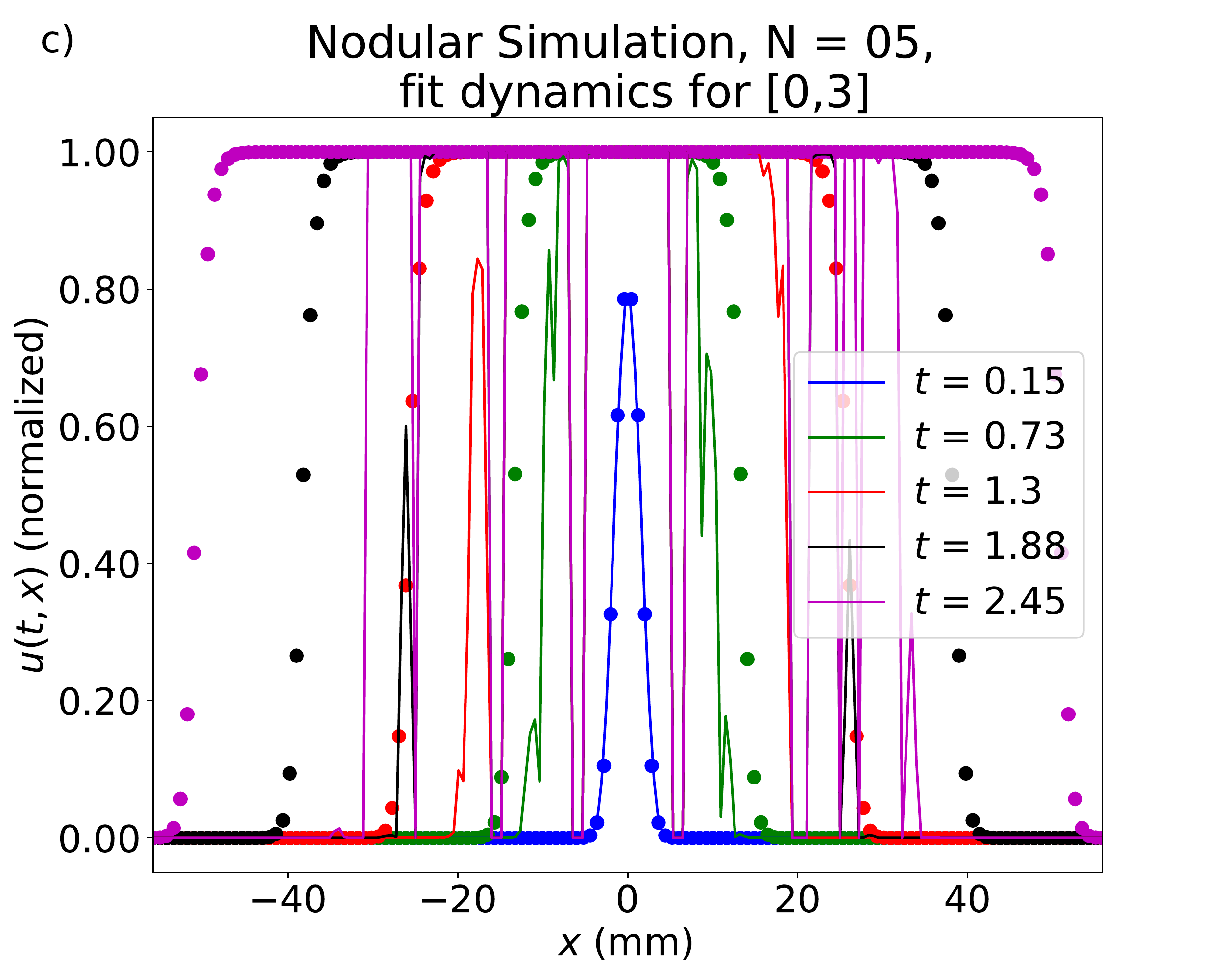}
    \includegraphics[width=0.45\textwidth]{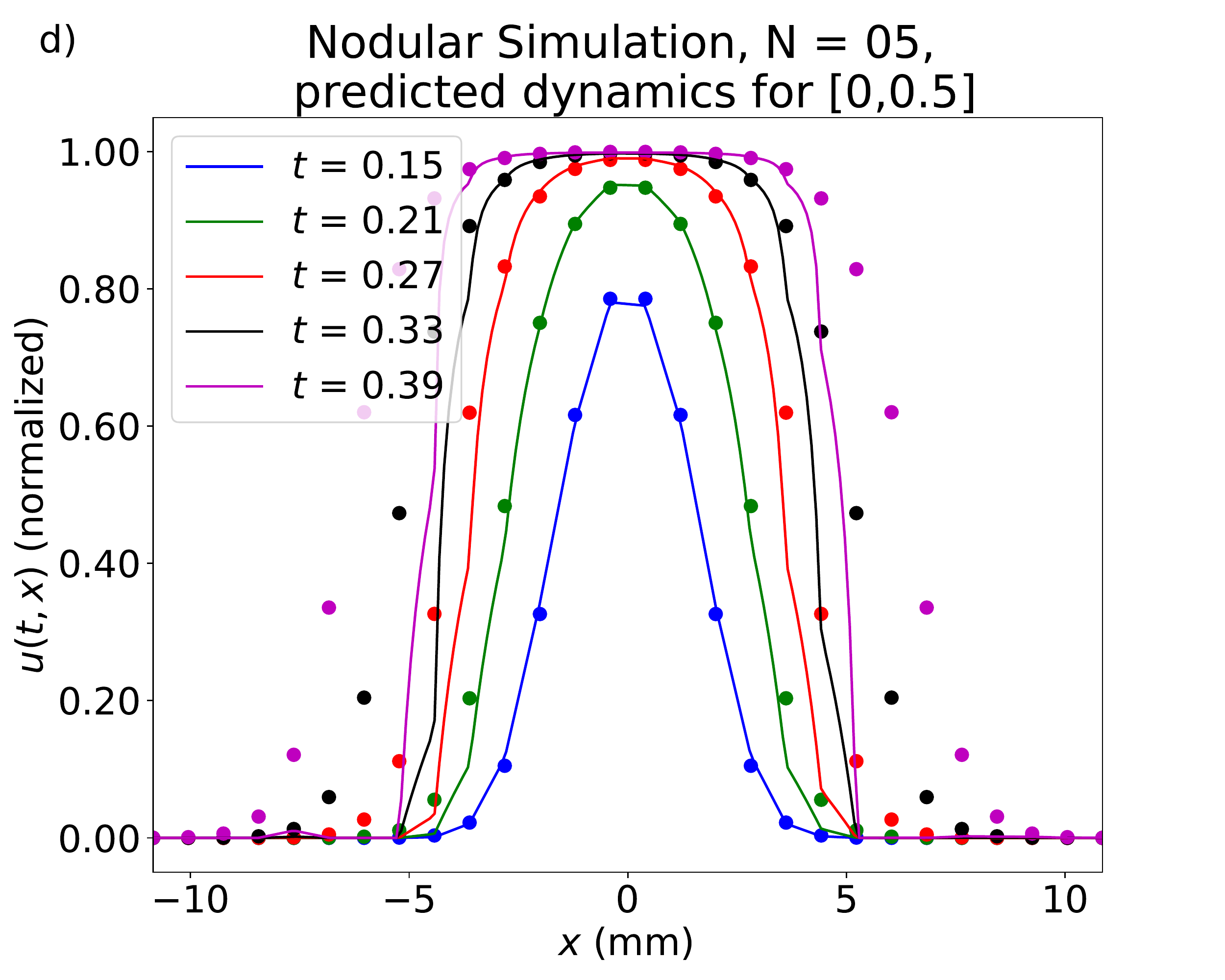}
    \caption{Fit and predicted dynamics for the nodular  with $N=5$ time samples and $1\%$ noise. (a) The simulated learned equation for the nodular  that was inferred from data sampled over the time interval [0,0.5]. (b) The model that was inferred over the time interval [0,0.5] is used to predict the dynamics over the time interval [0,3]. (c) The simulated learned Equation for the nodular  that was inferred from data sampled over the time interval [0,3]. (d) The model that was inferred over the time interval [0,3] is used to predict the dynamics over the time interval [0,0.5].
    While the simulations in part (c) may appear to be the result of an unstable numerical simulation, it instead is the result of a noisy initial condition combined with a an inferred ODE model of the form $u_t=-28.58u^2+28.55u$. Small bumps in the initial condition grow to confluence over time as depicted in this figure. }
    \label{fig:FKPP_generalization_3}
\end{figure}

\begin{figure}
    \centering
    \includegraphics[width=0.45\textwidth]{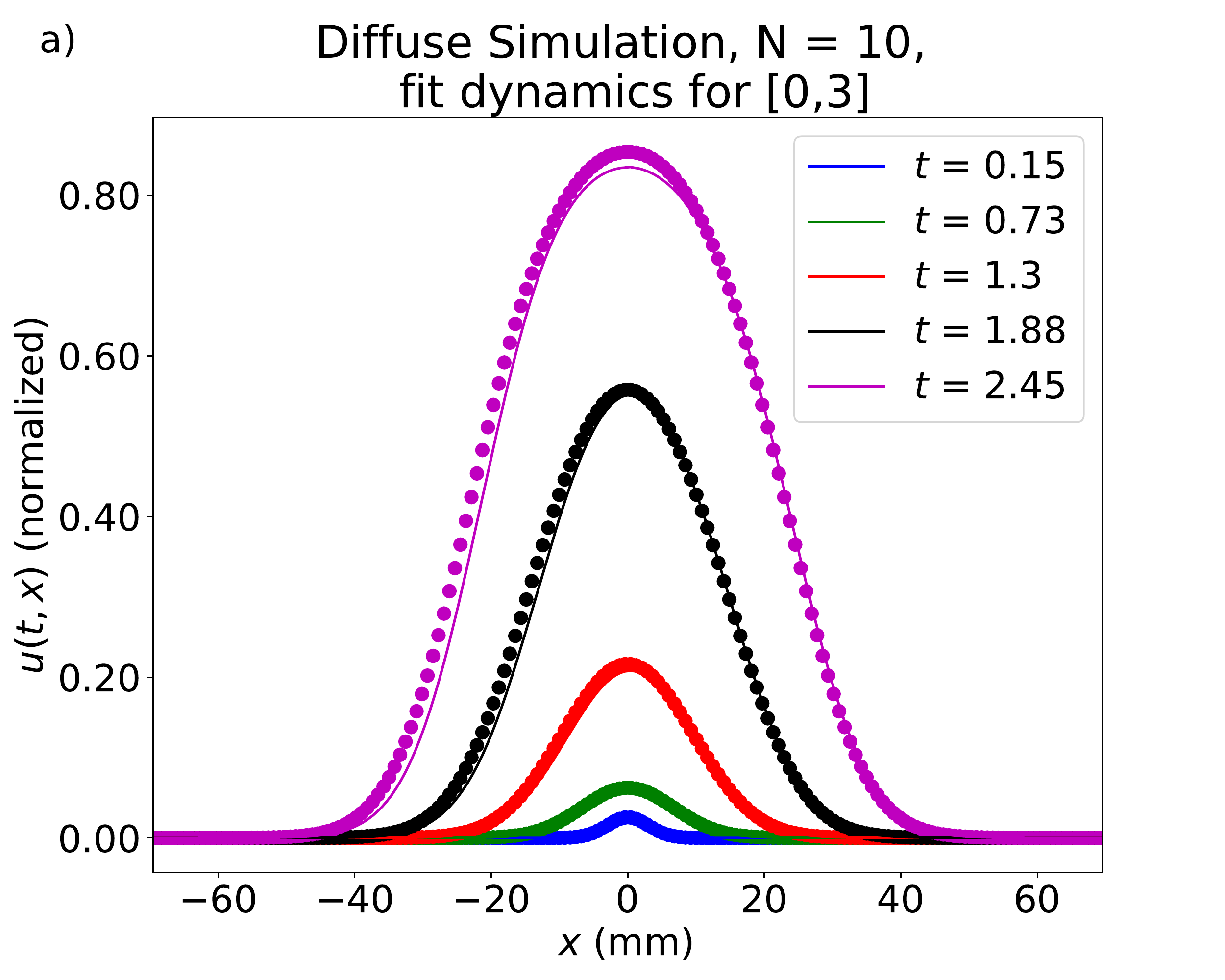}
    \includegraphics[width=0.45\textwidth]{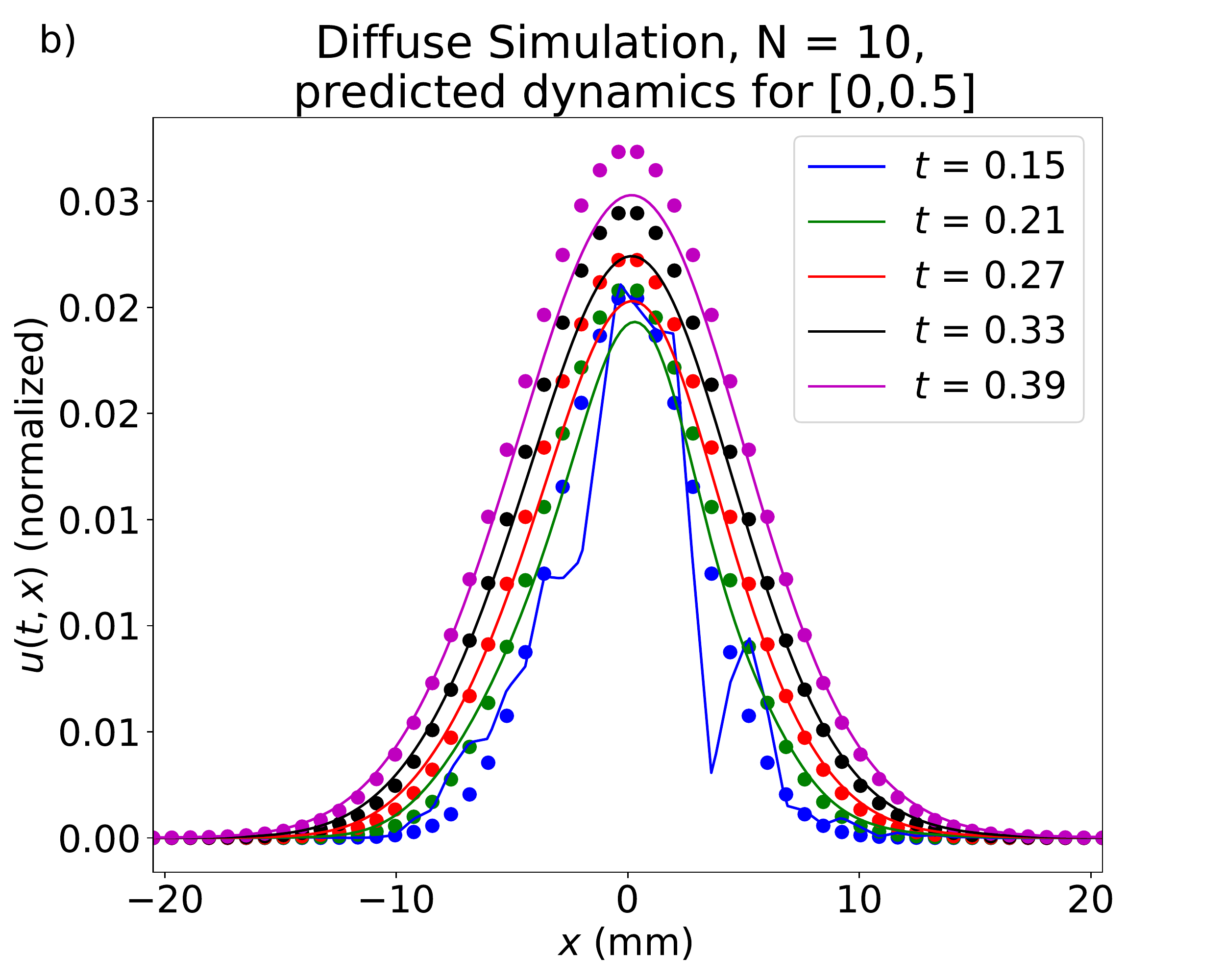}
    \includegraphics[width=0.45\textwidth]{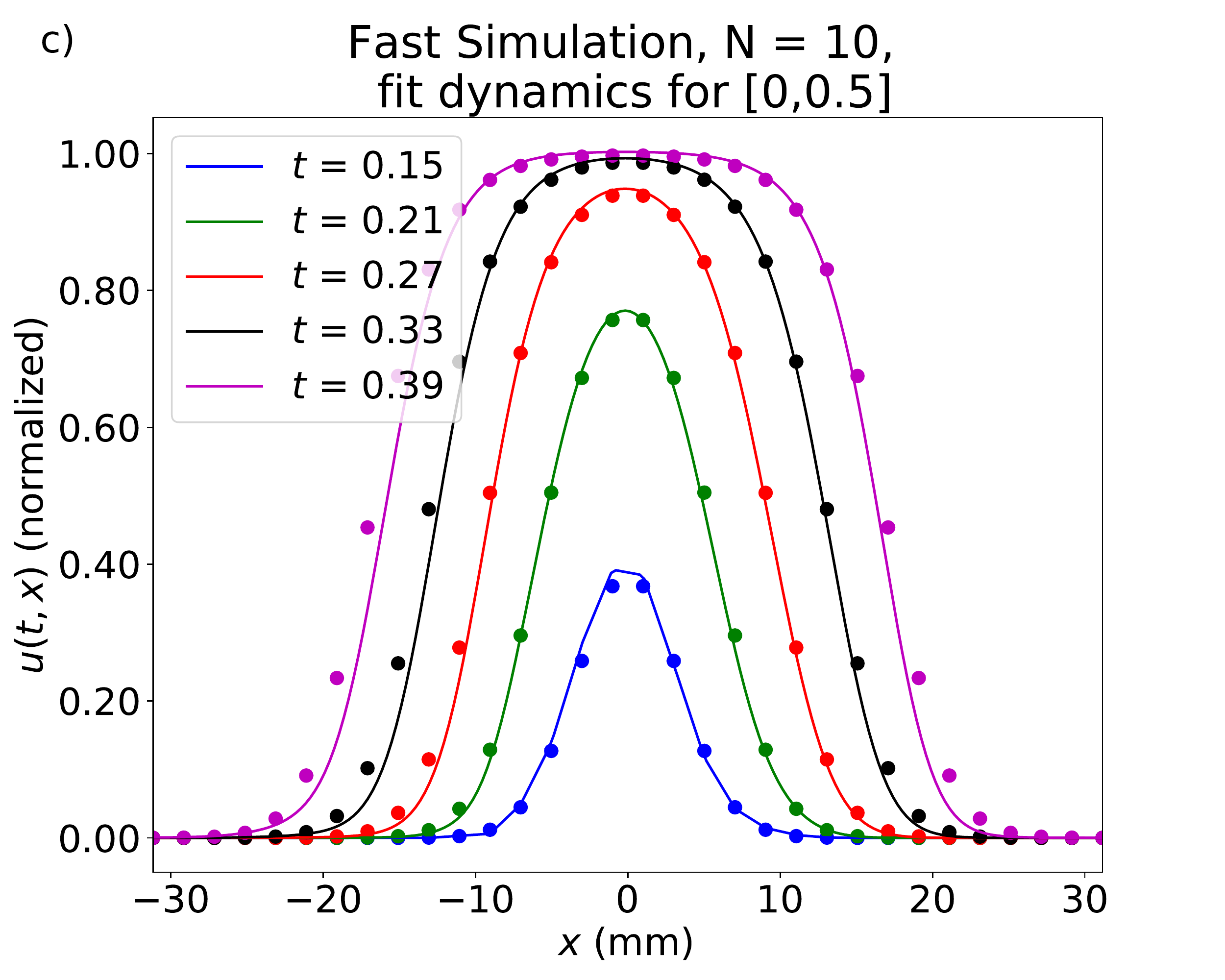}
    \includegraphics[width=0.45\textwidth]{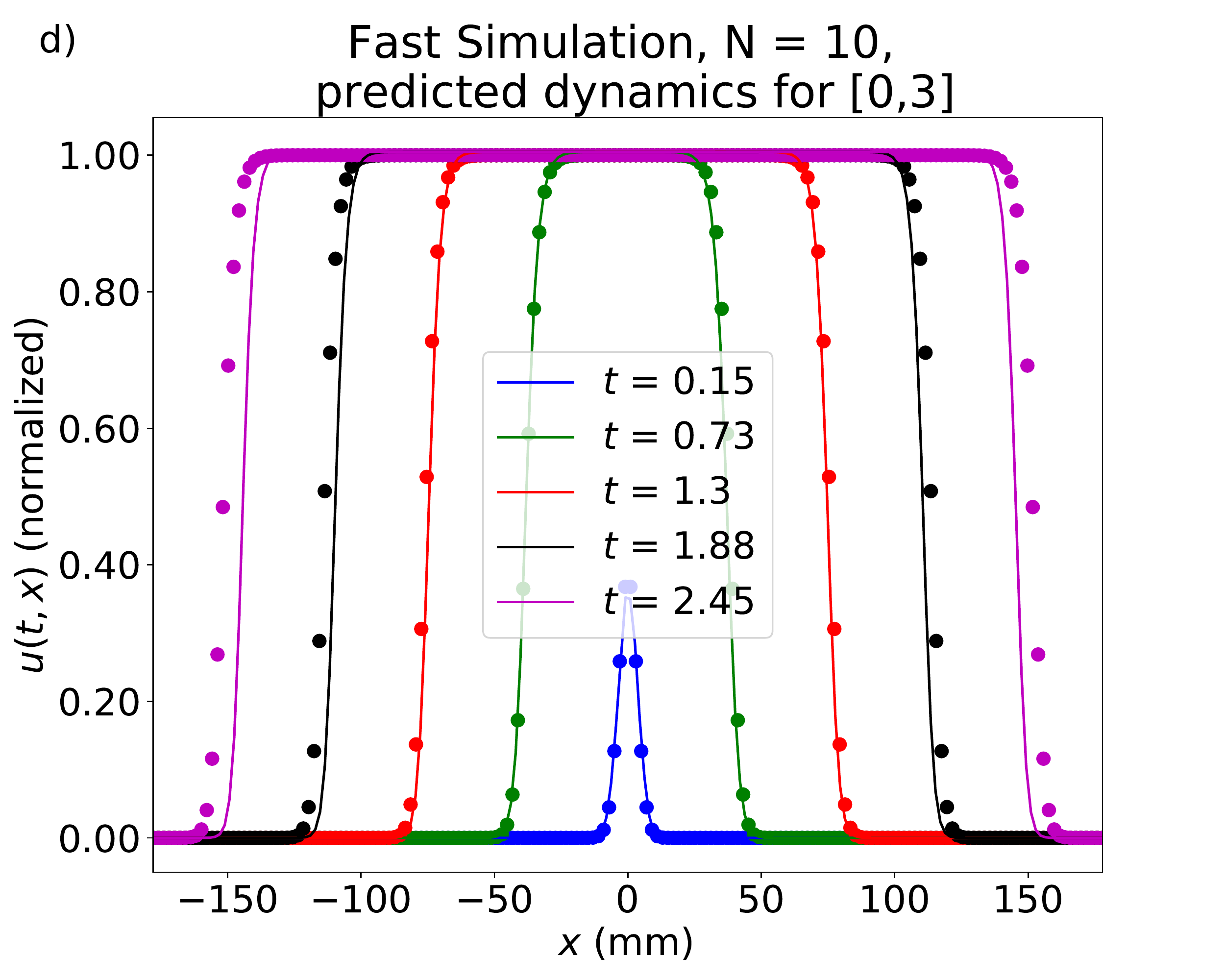}
    
    \caption{Sample fit and predicted dynamics for s with $N=10$ time samples and $5\%$ noise. (a) The simulated learned Equation for the diffuse  that was inferred from data sampled over the time interval [0,3]. (b) The model that was inferred over the time interval [0,3] is used to predict the dynamics over the time interval [0,0.5]. (c) The simulated learned Equation for the fast  that was inferred from data sampled over the time interval [0,0.5]. (d) The model that was inferred over the time interval [0,0.5] is used to predict the dynamics over the time interval [0,3].
    }
    \label{fig:FKPP_generalization_4}
\end{figure}

The fit and predicted system dynamics for the diffuse, fast, and nodular s with 1\% noise and $N=5$ time samples are depicted in Figures \ref{fig:FKPP_generalization_1}-\ref{fig:FKPP_generalization_3}, respectively.
The fit and predicted dynamics for the Diffuse and fast s with 5\% noise and $N=10$ time samples are depicted in Figures \ref{fig:FKPP_generalization_4}.

\clearpage
\bibliographystyle{spmpsci}      

\bibliography{references.bib}
\end{document}